\documentclass[twocolumn,aps,prx,showpacs,amsmath,superscriptaddress,longbibliography,notitlepage,floatfix]{revtex4-1}
\usepackage{amssymb}
\usepackage{mathrsfs}
\usepackage{graphicx}
\usepackage{braket}
\usepackage{float}
\usepackage[caption=false]{subfig}
\usepackage[pdftex,colorlinks,linkcolor={red!75!black},citecolor={red!75!black},urlcolor={blue!75!black}]{hyperref}
\usepackage{etoolbox}
\usepackage{csquotes}
\usepackage[normalem]{ulem}
\usepackage{epstopdf}
\usepackage{xcolor}
\usepackage{bm}
\usepackage{soul}

\renewcommand{\Im}{\operatorname{Im}}
\newcommand{\clr}{\color{red!75!black}}

\newcommand{\Rnum}[1]{\uppercase\expandafter{\romannumeral #1\relax}}
\def\LLH{\textcolor{cyan}}

\def\blue{\textcolor{black}}

\usepackage{soul}
\usepackage{changes}

\usepackage{verbatim}

\begin{document}

\title{Anyon-Induced Criticality and Dynamical Stability in Non-Hermitian Many-Body Systems
}
\author{Yi Qin}
\affiliation{Department of Physics, National University of Singapore, Singapore 117542}
\author{Yee Sin Ang}
\affiliation{Science, Mathematics and Technology, Singapore University of Technology and Design, Singapore 487372, Singapore}
\author{Linhu Li}\email{lilinhu@quantumsc.cn}
\affiliation{Quantum Science Center of Guangdong-Hong Kong-Macao Greater Bay Area (Guangdong), Shenzhen, China}
\author{Ching Hua Lee}\email{phylch@nus.edu.sg}
\affiliation{Department of Physics, National University of Singapore, Singapore 117542}

\begin{abstract}
We show that anyonic statistics fundamentally reshapes non-Hermitian many-body physics by intrinsically invalidating pseudo-Hermiticity, leading to a unique real-complex spectral transition with characteristically dense imaginary energies. 
This transition occurs exclusively under an anyonic exchanging phase,
revealing a form of non-Hermitian criticality driven purely by exchange statistics. 
\blue{The spectrum features a near-degenerate pair of dominant states, separated from the others by a much larger imaginary-energy gap, yielding anomalously stable short-time anyonic dynamics. Our results show that the occupation-dependent statistical string of anyons provides an effective nonlocal mechanism for generating unconventional non-Hermitian critical behaviors.
Entanglement further exhibits an anyon-specific exponential finite-size scaling, with a decay rate analytically determined by relative skin localization and no stable bosonic or pseudofermionic counterpart. The predicted dynamical stabilization is also verified on IBM quantum hardware.
}
\end{abstract}

\maketitle

\noindent \emph{{\clr Introduction}.---}
Abelian anyons, which interpolate continuously between bosons and fermions through a tunable statistical angle, have attracted longstanding interest as a fundamental extension of quantum statistics~\cite{Leinaas1977ICB,Goldin1980JMP,Goldin1981JMP,Wilczek1982PRL,Tsui1982PRL,Laughlin1983PRL,hao2008GPRA,hao2009GPRA,hao2012DPRA,Halperin1984PRL,Arovas1984PRL,Yao2007PRL,Bauer2014NC,kitaev2006A,Keilmann2011NC,Greschner2015PRL,Arcila2016PRA,Arcila2018PRA,Lange2017PRL,Liu2018PRL,Joyce2023arxiv}. They play an essential role in understanding fractional quantum Hall physics and quantum computation~\cite{kitaev2003fault,DasSarma2005PRL,Nayak2008RMP,Carrega2021,iqbal2024non,lee2023par,LeeChing2015,LeeChing2018PRL}. Recent experimental breakthroughs in ultracold atomic systems have successfully engineered Abelian anyons and observed their characteristic asymmetric spreading dynamics~\cite{dhar2025N,Joyce2023arxiv}, paving the way for controlled studies of exotic quantum statistics in synthetic matter.

In parallel, non-Hermitian physics has emerged as a powerful framework to describe open quantum systems with gain and loss, leading to unique phenomena absent in Hermitian settings. Among these, the non-Hermitian skin effect (NHSE)~\cite{Yao2018,Yao2018PRL,Martinez2018PRB,Lee2019,okuma2023non,zhang2022review,lin2023top,gohsrich2025non,yan2025hsg} — the 
{extensive} localization of all bulk states at system boundaries — has become a central topic of intense research for their unconventional responses \cite{qin2023kinked, huang2026complex, d5zcp1sk,zhao2025magne,shen2023proposal, z9m13mwb, XiaoLei2024PRL,wang2025nonlinear, wu2026observation,yang2025beyond,xue2025non,yang2025reversing,okuma2026ste,wang2025observation, yu2026sensitivity} in both single-particle~\cite{Nobuhiro2026, lin2026glo, rahul2026cont, saito2026quas,  bai2026eng,  s26b8bdl, WuPRL2025,li2025exact,li2022non,   liu2025anom, li2025algebraic,rafi2022unconventional, hu2025topological, ammari2025non, han2026observation, li2025anderson,  Zhang2024PRB, ZhangPRL2023, bmq57tf6, yang2025conf, 4yt24rx4,  lei2025inter,    SCPHe2023,cwwdbclc,3927n25r,li2025phase} and many-body contexts~\cite{bhpz17d2,hao2025interacting,yang2025non, deng2026conf, yi2026dir, longhi2026erra,koh2025interacting,long2026sym}.
{More recently, a new form of criticality across the NHSE-deformed complex momentum solutions -- the critical skin effect (CSE) -- has been identified as a source of
system size-dependent scale-free localization, fundamentally rooted in the competition between multiple NHSE localization channels}~\cite{Li2020NC, Liu2021, rafi2022CSE, liang2025size,rafi2025critical,yang2024percolation,ZhangPRB2025,liu2025non,meng2025gen,xu2025exciton,cheng2025stochasticity,cai2024non,xue2026topologically}.

In the many-body regime, {the} 
NHSE has led to even more sophisticated emergent phenomena~\cite{Lee2021RS,Faugno2022PRL,Zheng2024PRL,Mu2022,Cao2023PRB,Louis2024SCP,MaoLiang2023,ZhangPRB2022,LeeC2021PRB,Xingran2021PRB,yoshida2023non,hamanaka2024,gliozzi2024many,QinYi2024PRL,qin2025CP,roccati2026p},
such as Fermi-like surfaces in real space and bosonic boundary condensation~\cite{Mu2022,Cao2023PRB,yang2025non,Louis2024SCP,shen2025observation}, with the latter being strongly suppressed by repulsive interactions~\cite{MaoLiang2023,Zheng2024PRL}. 
{Rich classes of new many-body CSE states have also been identified in multi-component bosonic systems}~\cite{qin2025many}. \blue{Anyonic many-body systems are closely connected to conformal field theories and quantum-group structures, particularly in interacting anyon chains and integrable lattice models
\cite{PasquierSaleur1990,Feiguin2007GoldenChain,Trebst2008FibonacciAnyons}.
Non-unitary topological theories and lattice realizations may also involve non-Hermitian Hamiltonians
\cite{Freedman2012GaloisConjugates}, while non-Hermitian quantum mechanics has long been linked to statistical-mechanical mappings, $\mathcal{PT}$ symmetry, and, more recently, conformal criticality~\cite{HN1996PRL,BenderPRL1998,Chou2025PT}.
However, the direct role of anyonic exchange phases in non-Hermitian spectral and boundary phenomena, particularly the NHSE with effective nonlocal couplings, remains comparatively underexplored.}


In this Letter, we reveal a new type of critical phenomena fundamentally rooted in the anyonic statistics of coupled Abelian anyonic systems. {Namely, through the nonlocality in the basis space of quasi-one-dimensional (1D) systems,
anyonic statistics explictly breaks the pseudo-Hermiticity and substantially alters the spectral structure},
{so to allow an infinitesimal hopping perturbation to} induce a real-complex spectral transition in the thermodynamic limit. In contrast to many-body CSE that relies on the order of hopping processes between scattering and/or bound states, such anyon-induced spectral transitions (AISTs) remains robust across different mixtures of these states, unveiling criticality insensitive to the explicit particle configurations.
Furthermore, the anyonic statistics also {reshape the band structure} of the imaginary spectrum, leading to an {experimentally salient} enhanced stability of the short-time quench dynamics.
\begin{figure}[htbp]
\centering
\includegraphics[height=4.1cm]{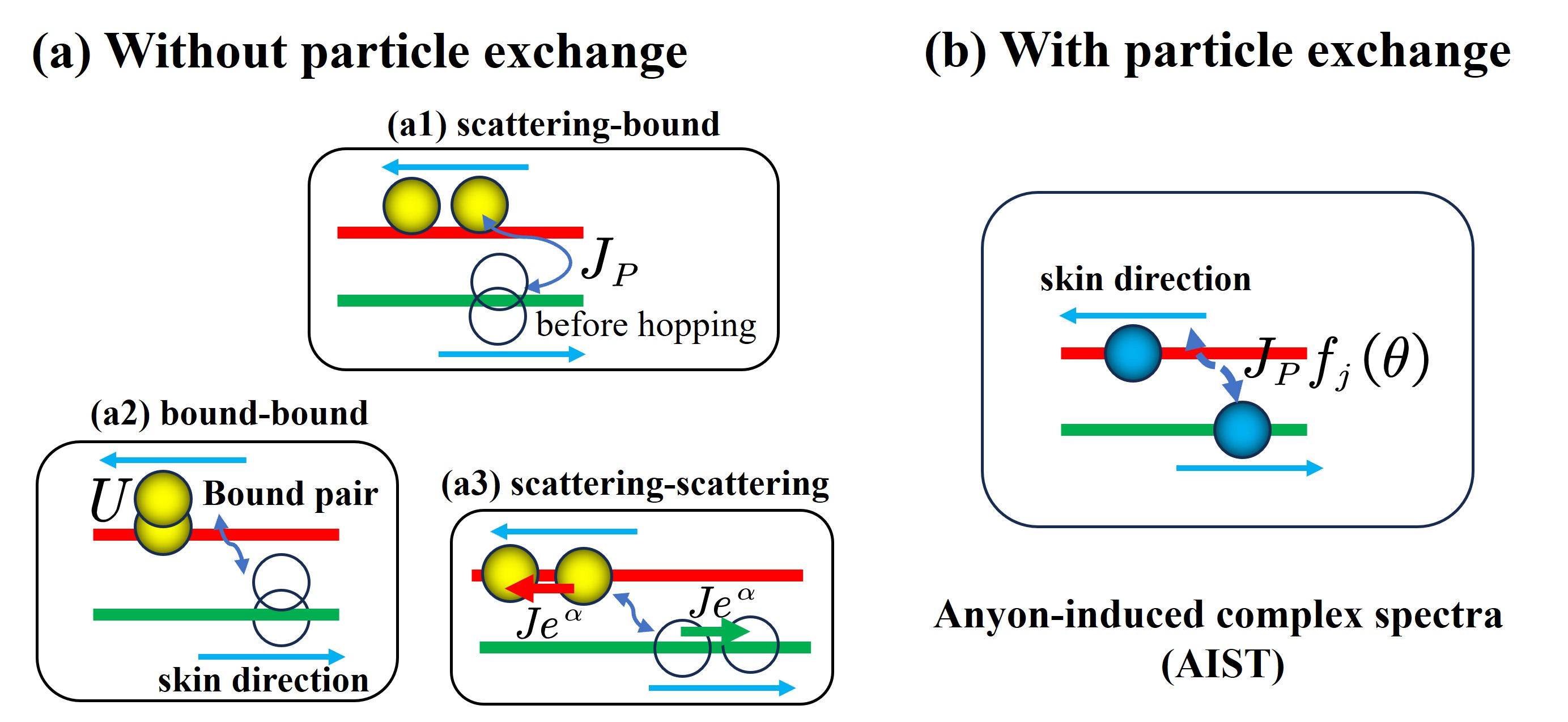}
\caption{\label{Schematic}  Schematic illustration of representative interaction processes underlying the AIST
in coupled Abelian anyonic systems. 
(a) Processes without particle exchange: (a1) a scattering-bound process hybridizing extended and bound two-particle configurations, (a2) a bound-bound process renormalizing the bound-state sector, and (a3) a scattering-scattering process reshaping the scattering continuum. These channels are determined by the composition of scattering and bound states and do not explicitly depend on the statistical angle $\theta$. 
(b) 
A particle-exchange process, where
interchain hopping acquires an occupation-dependent phase term
$f_j(\theta)$. This process
introduces a $\theta$-dependent asymmetry, breaks pseudo-Hermiticity, and leads to the AIST.
}\end{figure}

\noindent \emph{{\clr Model for interacting 1D anyons}.---}
Our study focuses on 
1D abelian anyons satisfying the commutation relations
\blue{
\begin{equation}
\begin{array}{l}
{[{{\hat a}_j},{{\hat a}_k}]_\theta } \equiv {{\hat a}_j}{{\hat a}_k} - {e^{ i\theta {\mathop{\rm sgn}} (j - k)}}{{\hat a}_k}{{\hat a}_j} = 0,\\
{[{{\hat a}_j},\hat a_k^\dag ]_{ - \theta }} \equiv {{\hat a}_j}\hat a_k^\dag  - {e^{-i\theta {\mathop{\rm sgn}} (j - k)}}\hat a_k^\dag {{\hat a}_j} = {\delta _{jk}}.\label{commute}
\end{array}
\end{equation}
}
$\hat a_j^\dag$ is the particle creation operator, describing bosons when $\theta=0$ and pseudo-fermions when $\theta=\pi$~\cite{Liu2018PRL,Keilmann2011NC}.
To generate interacting skin channels and effective nonlocal hoppings, we consider a quasi-1D ladder with two sublattices and onsite interaction, described by the Hamiltonian
$
\hat H=\sum_{\sigma=A,B}\hat H_\sigma+\hat H_{AB}+\hat H_{\rm onsite}
$,
with 
\begin{small}
\begin{align}
\hat H_\sigma &=
-J\sum_{j=1}^{L-1}\left(
e^{\alpha_\sigma}\hat a_{j,\sigma}^\dagger \hat a_{j+1,\sigma}
+e^{-\alpha_\sigma}\hat a_{j+1,\sigma}^\dagger \hat a_{j,\sigma}
\right), \label{eq:H_sigma}
\\[-3pt]
\hat H_{AB} &=
J_p\sum_{j=1}^L\left(
\hat a_{j,A}^\dagger \hat a_{j,B}
+\hat a_{j,B}^\dagger \hat a_{j,A}
\right),
\\[-3pt]
\hat H_{\rm onsite} &=
\frac{U}{2}\sum_{j=1,\sigma}^L
\hat n_{j,\sigma}\left(\hat n_{j,\sigma}-1\right)
+\mu\sum_{j=1}^L\left(
\hat n_{j,A}-\hat n_{j,B}
\right).\label{eq:H_full}
\end{align}
\end{small}
Here \(\sigma=A,B\) labels the two sublattices. The asymmetric hopping
amplitudes \(J e^{\pm\alpha_\sigma}\) introduce non-reciprocity, with
\(\alpha_{A/B}=\pm\alpha\), while $J_p$ describes inter-sublattice hybridization. \blue{These hoppings define the quasi-1D ladder geometry, while all microscopic hopping processes remain spatially short-ranged.} The onsite term \(\hat H_{\rm onsite}\)
includes the Hubbard interaction \(U\) and a staggered chemical potential
\(\pm\mu\) with
\(\hat n_{j,\sigma}=\hat a_{j,\sigma}^\dagger\hat a_{j,\sigma}\).

Using a generalized Jordan--Wigner transformation
\begin{equation}
\hat a_j=\hat b_j \exp\!\left(-i\theta\sum_{k=1}^{j-1}\hat n_k\right)
\end{equation}
with the ladder sites $\alpha_j$ ordered as $(A_1,...,A_L,B_1,...B_L)$
({see Supplemental Sec.~\ref{sec:sm-jw}})~\cite{Liu2018PRL, qin2025CP, Bonk2025PRL},
the anyonic Hamiltonian is mapped onto a bosonic Hamiltonian
$
\hat H_{\rm eff}=\sum_{\sigma=A,B}\hat H_\sigma^{\rm eff}
+\hat H_{AB}^{\rm eff}
+\hat H_{\rm onsite}
$.
It takes the same form as in Eqs.~\eqref{eq:H_sigma} to \eqref{eq:H_full}, only with hopping terms acquiring occupation-dependent Peierls phases,
\begin{align}
\hat{a}_{j,\sigma}^\dagger\hat{a}_{j+1,\sigma}&\rightarrow \hat{b}_j^\dagger\hat{b}_{j+1}e^{-i\theta\hat{n}_{j,\sigma}},
\hat{a}_{j,B}^\dagger\hat{a}_{j,A}&\rightarrow
f_j(\theta)
\hat b_{j,B}^\dagger \hat b_{j,A}\nonumber
\end{align}
and their Hermitian conjugate, with $f_j(\theta)=e^{i\theta\left(
\sum_{k=j}^L \hat n_{k,A}
+\sum_{k=1}^{j-1}\hat n_{k,B}
\right)}$. Here the number operator satisfies
\(\hat n_{j,\sigma}=\hat a_{j,\sigma}^\dagger\hat a_{j,\sigma}
=\hat b_{j,\sigma}^\dagger\hat b_{j,\sigma}\),
and {importantly the statistical angle $\theta$} is encoded in  \(\hat H_\sigma^{\rm eff}\)  and  \(\hat H_{AB}^{\rm eff}\) as local and nonlocal hopping phases, respectively. \blue{Note that the transformation requires careful treatment under different boundary conditions (see the Supplemental Material Sec.~\ref{sec:sm-jw} for details). In this work, we focus exclusively on open boundary conditions.}

\begin{figure*}[htbp]
\centering
\includegraphics[height=7.6cm]{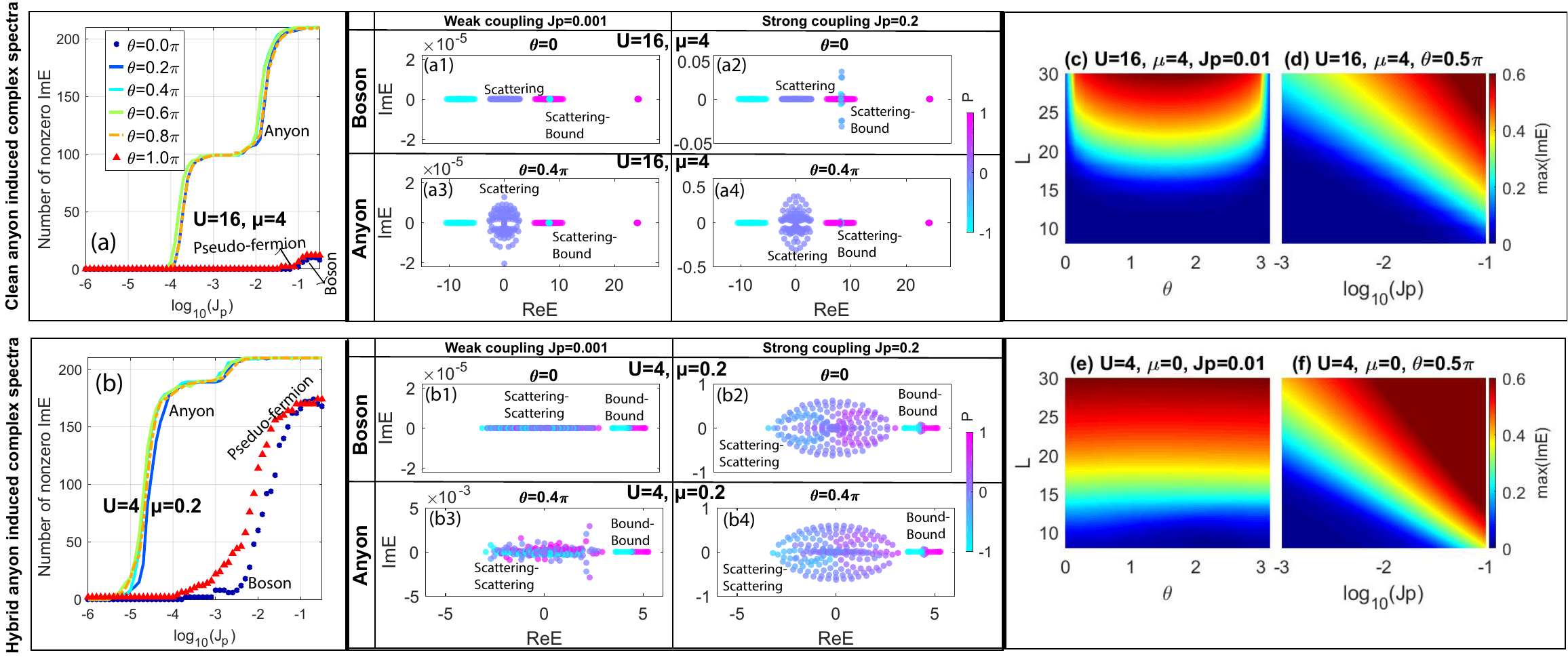}
\caption{\label{N0ReEvsJp} 
\textbf{Clean (Top Row) and Hybrid (Bottom Row) AIST for two particles.} 
Anyons, unlike bosons ($\theta=0$) and pseudofermions ($\theta=\pi$), generically undergo a real-complex transition in the scattering subspace at small $J_p$. 
(a) Number of nonzero imaginary energies (\(\mathrm{Im}E>10^{-7}\) in numerics) as a function of \(J_p\), with \(U=16,\ \mu=4\).
Compared with bosons and pseudofermions,
anyons acquire complex eigenenergies at an exponentially smaller $J_p$.
(a1) to (a4) show representative spectra colored by the sublattice polarization
\(P=(N_A-N_B)/(N_A+N_B)\), with \(N_\sigma=\sum_x\langle \hat{n}_{x,\sigma}\rangle\).
The mixture of scattering and bound states lead to the mixed many-body CSE that generates complex eigenenergies under a weak $J_p$ [(a1) and (a2)],
which are dominated by those from AIST in both their numbers and magnitudes [(a3) and (a4)]. (b) and (b1) to (b4) show the same quantities as (a) and (a1) to (a4), but with \(U=4,~\mu=0.2\).
Scattering and bound states are now separated in real energy.
The resultant AIST and many-body CSE are qualitatively similar, except that the scatter many-body CSE overwhelms the AIST under a relatively strong $J_p$ in (b4).
(c) and (d) Maximum imaginary eigenenergy, Max(Im$E$), for $U=16$ and $\mu=4$ as in (a).
In (c), AIST is clearly recognized by nonzero Max(Im$E$) only with \(\theta\neq0\) or \(\pi\), respectively.
Its critical nature can be identified in (d) by the threshold of $J_p$ to induce nonzero Max(Im$E$), which approaches zero exponentially as the system size $L$ increases.
(e) and (f) the same as (c) and (d), but with $U=4$ and $\mu=0$.
The scattering many-body CSE dominates the complex eigenvalues, leading similar Max(Im$E$) for arbitrary $\theta$ values in (e). 
In (f), the threshold of $J_p$ also decreases exponentially as $L$ increases.
Other parameters are \(J=e^{\alpha}=1/\sqrt{2}\) and \(L=10\), unless otherwise specified.
}
\end{figure*}

\noindent \emph{{\clr AIST and criticality induced by anyonic statistics}.---}
{A clean instance of the AIST is reported}
in Fig.~\ref{N0ReEvsJp}(a), 
displaying the numbers of eigenenergies with nonzero imaginary parts ($|{\rm Im}E|>10^{-7}$ in numerics)  as {$\theta$ varies}. 
{Saliently, for anyons ($\theta\neq0$ or $\pi$), exponentially smaller inter-sublattice coupling $J_p$ can already induce complex eigenenergies and thus amplification.}
Interestingly, anyons exhibit two onsets of complex spectra [the double plateaux in Fig.~\ref{N0ReEvsJp}(a)] due to the coexistence of different orders of hopping processes between Fock states (see Supplemental Material Sects.~\ref{sec:sm-perturbation} and \ref{sec:sm-pseudohermiticity}).
In the following we shall focus only on the first transition. 

To understand the origin of AIST,
 note that the Hamiltonian terms $\hat{H}_\sigma$ and $\hat{H}_{\rm onsite}$ satisfy a pseudo-Hermitian symmetry that ensures real eigenenergies in its unbroken phase  ({see supplemental Sec.~\ref{sec:sm-pseudohermiticity}}),
 \begin{equation}
\mathcal{K} (\hat{H}^{\rm eff}_\sigma+\hat{H}_{\rm onsite})\mathcal{K}^\dag = (\hat{H}^{\rm eff}_\sigma+\hat{H}_{\rm onsite})^\dag,
\end{equation}
with $\mathcal{K} = \mathcal{R}_z \mathcal{I} \mathcal{T}$, 
$\mathcal{I}$ the spatial inversion operator, 
$\mathcal{T}$ the time-reversal operator, and
$\mathcal{R}_z = \exp\left(-i \theta \sum_j \sum_{\sigma} \hat{n}_{j,\sigma}(\hat{n}_{j,\sigma}-1)/2\right)$ a density-dependent phase rotation.
However, the inter-sublattice term
$\hat{H}_{AB}^{\rm eff}$ {breaks} the $\mathcal{K}$-symmetry,
\begin{equation}
\mathcal{K}\hat{H}^{\rm eff} _{AB}\mathcal{K}^\dagger \neq (\hat{H}^{\rm eff} _{AB})^\dagger,
\end{equation}
leading to nonzero Im$E$ when $J_p$ cannot be ignored.
The only exceptions are $\theta=0$ and $\pi$, where the pseudo-Hermitian symmetry is restored for the full Hamiltonian,  eliminating the AIST {unique to} anyonic statistics.

Consistent with the above analysis, in Fig. \ref{N0ReEvsJp}(a), the system with $\theta=0$ or $\pi$ supports no complex eigenenergies until $J_p\gtrsim 10^{-2}$,
which indicates a symmetry-breaking scenario, where many-body CSEs emerge from interference between different many-body states [Fig.~\ref{Schematic}(a)].
Specifically for the chosen parameters, a mixed many-body {bosonic} CSE emerges between scattering and bound states mixed in real energy, as shown in Figs. \ref{N0ReEvsJp}(a1) and \ref{N0ReEvsJp}(a2).
In contrast, anyonic statistic invalidates the $\mathcal{K}$-symmetry. Consequently, complex eigenergies appear at an exponentially smaller  $J_p$,
and dominates
in both their numbers and  imaginary magnitudes compared to the mixed many-body CSE, as can be seen from Figs. \ref{N0ReEvsJp}(a3) and \ref{N0ReEvsJp}(a4).

Another example is demonstrated in Figs. \ref{N0ReEvsJp}(b),
where the system supports many-body CSEs from the interference between the same type of energy clusters [bound or scattering ones, see Figs. \ref{N0ReEvsJp}(b1) and \ref{N0ReEvsJp}(b2) for an example for bosons]. 
Similarly to the previous case [Figs. \ref{N0ReEvsJp}(a)], 
anyonic statistics enhances the sensitivity and generate more complex eigenenergies at a smaller $J_p$.
Nevertheless, the scattering many-body CSE involves hopping processes only between scattering states, thus it is much more significant than the mixed many-body CSE that breaks interaction-bound pairs to form scattering states, and may overwhelm the AIST under relatively stronger $J_p$, as shown in Fig. \ref{N0ReEvsJp}(b4).

To clarify the AIST and many-body CSE mechanisms,
in Figs.~\ref{N0ReEvsJp}(c) to ~\ref{N0ReEvsJp}(f) we demonstrate the {maximal} Im$E$ versus $\theta$, $J_p$, and $L$, for the previous two representative examples.
In Fig.~\ref{N0ReEvsJp}(c) with larger \(\mu\) and \(U\), the maximum Im$E$ jumps from zero to a finite value as \(\theta\) deviates from \(0\) and \(\pi\), signaling a statistics-driven criticality from the AIST.
In contrast,  scattering many-body CSE in the the regime with weaker  \(U\) and \(\mu\) becomes the main origin of complex eigenenergies, making the maximum Im$E$ insensitive to $\theta$,
as shown in Fig.~\ref{N0ReEvsJp}(e).
Nevertheless, as seen in Figs.~\ref{N0ReEvsJp}(d) and \ref{N0ReEvsJp}(f), 
the threshold of $J_p$ to induce complex eigenenergies in both cases tends to zero as $L$ approaches infinity, unveiling qualitatively the same criticality for many-body CSE and AIST, despite that they origin from different symmetry-breaking or -invalidation mechanisms. \blue{An independent averaged entanglement diagnostic further
reveals an analytically-determined exponential scaling unique for anyons
(see End Matter).}

\begin{figure}[htbp]
\centering
\includegraphics[height=8cm]{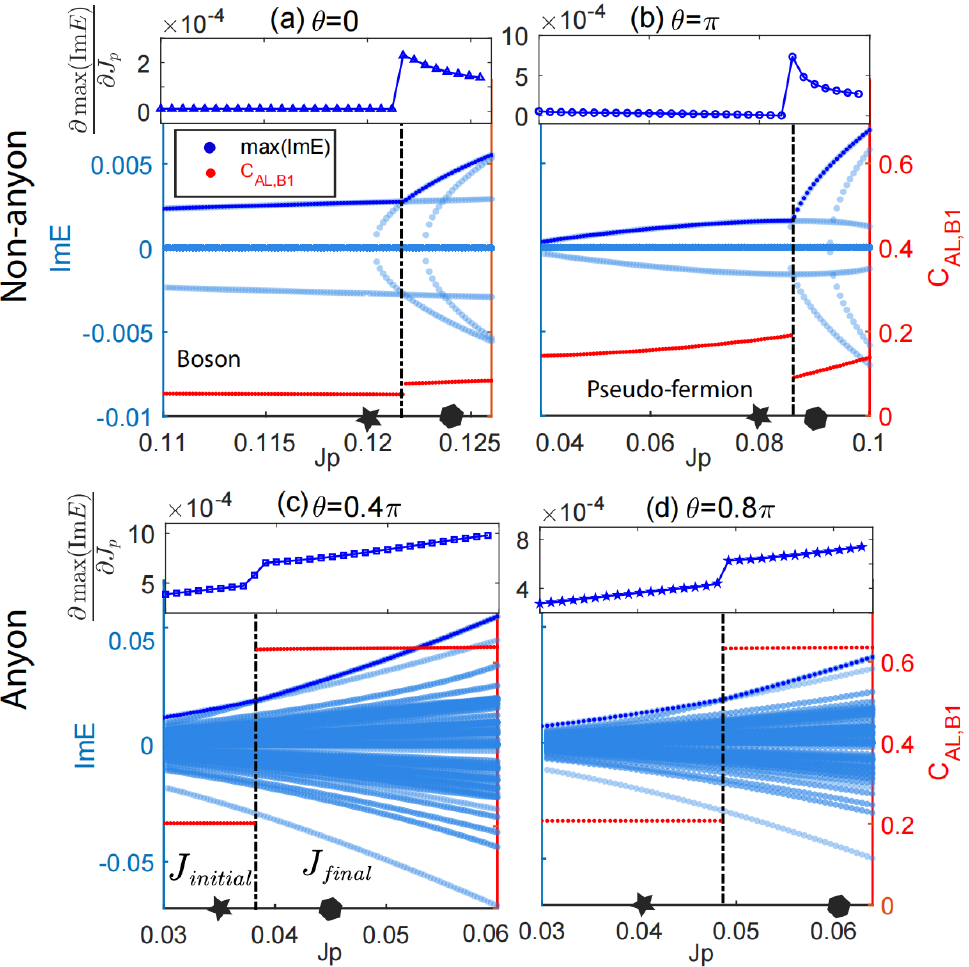}
\caption{\label{anyonPT} \textbf{Distinct imaginary density of states and spectral transitions for anyons and non-anyons.} 
Anyons exhibit broadly complex spectra, indicating pseudo-Hermiticity invalidation via AIST.
(a) and (b) imaginary spectrum (blue) versus $J_p$ for bosons and pseudofermions, respectively. 
Discontinuous jumps between real and complex eigenenergies signal the pseudo-Hermiticity breaking in the many-body CSE.
A transition with Max(Im$E$) (dark blue) swapping between two states is clearly seen in either case (vertical dashed line).
Top panels show the derivative of Max(Im$E$), which jumps discontinuously at the transition point.
(c) and (d) the same as (a) and (b), but for anyons with different $\theta$.
Eigenenergies are generally complex, indicating the pseudo-Hermiticity invalidation of AIST.
\blue{The spectrum exhibits an increased proportion of imaginary-energy eigenvalues}, yet the transition of Max(Im$E$) still persists, as shown by the discontinuous change of its derivative.
The transition is also marked by the jump of the edge correlation $C_{A_LB_1}$ of the maximum-Im$E$ state (red dots). 
The imaginary energy distribution and its response to coupling differ markedly between anyons and non-anyons.
Other parameters are \( U = 16 \), \( \mu = 4 \), and, \( L = 10, J=e^{\alpha}=1/\sqrt{2} \). 
black stars and hexagons mark the parameters before and after quench is Fig. \ref{fig:Quench}. 
} 
\end{figure}

\noindent \emph{{\clr Anyon-enhanced imaginary spectral density of states}.---} \blue{The statistics-induced nonlocal dressing of non-Hermitian anyonic interactions also manifests as a more prominent imaginary-energy component of the spectrum}, as shown in Fig.~\ref{anyonPT}. For bosons and pseudofermions [Figs.~\ref{anyonPT}(a) and~\ref{anyonPT}(b)], {most eigenenergies remains real, with a few acquiring} nonzero Im$E$ as the coupling $J_p$ increases. Notably, these imaginary energies cross each other at certain $J_p$, swapping the maximum-Im$E$ state across the crossing.
{Such distinct swapping of maximal Im$E$ results in a sharp discontinuity in the first derivative} (top of each panel). In contrast, occupation-dependent statistical-string dressing from the anyonic phase $\theta$ gives macroscopically many eigenstates with sizable Im$E$, with the maximum-Im$E$ state forming an isolated near-degenerate pair [Figs.~\ref{anyonPT}(c) and~\ref{anyonPT}(d)].
The swapping of maximal Im$E$ is comparably less clear, but can also be identified by a slightly distinct discontinuity in the derivative.

Note that imaginary eigenenergies arise from the non-negligible effect of $J_p$, which couples the two sublattices $A$ and $B$ with rightward and leftward non-reciprocal pumping, respectively.
{This tendency is captured by the } boundary correlation function
\begin{equation}
C_{A_LB_1}=\langle \phi |\hat{n}_{A,L}\hat{n}_{B,1}|\phi \rangle
\label{eq:CLA1B}
\end{equation}
which measures the density–density correlation between site $A_L$ and $B_1$
for the maximum-Im$E$ state $|\phi\rangle$.
As shown in the {red dashed curves in} Fig.~\ref{anyonPT}, $C_{A_LB_1}$ jumps abruptly when the maximum Im$E$ state switches. Surprisingly, despite a less noticeable change in Im$E$ gradient for anyons, their
$C_{A_LB_1}$ jumps more drastically, suggestive of non-local state transitions beyond what is perceptible through spectral transitions alone.

\begin{figure}[t]
\centering
\includegraphics[width=\linewidth]{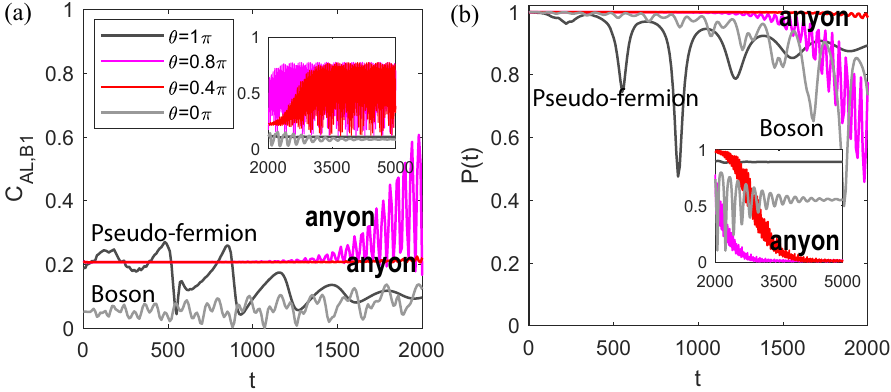}
\caption{\label{fig:Quench} \textbf{Dynamical stability induced by anyonic statistics.}
(a) and (b) boundary correlation $C_{A_LB_1}(t)$ and the overlap between final and initial states $P(t)$,
respectively.
The pre- and post-quenched parameters are $(J_{\mathrm{ini}}, J_{\mathrm{{final}}})=(0.12, 0.124)$ for $\theta=0$, 
$(0.035, 0.045)$ for $0.4\pi$, 
$(0.04, 0.06)$ for $0.8\pi$, 
and $(0.08, 0.09)$ for $\pi$, as marked by black stars and hexagons in Fig.~\ref{anyonPT}.
In the short-time regime with $t\lesssim10^3$, bosons and pseudofermions show strong fluctuations, while anyons are dynamically stable after quench{ing}.
\blue{Beyond this regime, anyons exhibit {strong oscillations} in $C_{A_LB_1}(t)$ because the two dominant states are nearly degenerate in Im$E$} ({see supplemental Sec.~\ref{sec:sm-eigenenergies}}). Other parameters are $N=2$, $U=16$, $\mu=4$, $L=10$, and $J=e^{\alpha}=1/\sqrt{2}$. }
\end{figure}

\noindent \emph{{\clr Anyon-induced stabilization of quench dynamics}.---} The proliferation of  imaginary eigenenergies for anyons also leads to dynamical consequences. Here
we consider a quench across two coupling parameter values $J_{\mathrm{ini}}$ and $J_{\mathrm{{final}}}$ chosen across the maximum-Im$E$ state transitions in Fig.~\ref{anyonPT}, marked by the black stars and hexagons:  
\begin{equation}
\begin{cases}
t=0: & J_p=J_{\mathrm{ini}},\\
t>0: & J_p=J_{\mathrm{{final}}},
\end{cases}
\end{equation}
where the system is initialized in the maximum-Im$E$ eigenstate $|\psi_{\rm ini}\rangle$ of $H(J_{\mathrm{ini}})$ and then {evolving} as
\begin{equation}
|\psi(t)\rangle =
\frac{e^{-iH(J_{\mathrm{{final}}})t}|\psi_{\rm ini}\rangle}
{\sqrt{\langle\psi_{\rm ini}|\,e^{+iH^\dagger(J_{\mathrm{{final}}})t}\,e^{-iH(J_{\mathrm{{final}}})t}\,|\psi_{\rm ini}\rangle}}\,.
\end{equation}
In Fig.~\ref{fig:Quench}(a), anyons are seen to have very stable 
boundary correlation in short-time dynamics, suggesting the domination of the  pre-quench state $|\psi_{\rm ini}\rangle$ due to the anyonic spectral feature. 
\blue{That is, by ordering  the eigenenergies as
$\Im E_0\geq \Im E_1\geq \Im E_2\geq\cdots$ and defining
$\delta_{01}=\Im E_0-\Im E_1$ and
$\Delta_{\rm sep}=\Im E_1-\Im E_2$,
we obtain $\delta_{01}\sim10^{-5}$ and $\Delta_{\rm sep}/\delta_{01}\sim10^2$ for the post-quench anyonic spectrum, making the two leading states a near-degenerate pair well separated from the remaining spectrum. (Supplemental Material, Sec.~\ref{sec:sm_imag_spacing}).}

\blue{The dynamically stable regime can be estimated using a two-state approximation,
$|\psi(t)\rangle\simeq c_2e^{-iE_2t}|\psi_2\rangle+c_ke^{-iE_kt}|\psi_k\rangle$
($k=0,1$), giving
$|c_k(t)/c_2(t)|=|c_k/c_2|e^{(\Im E_k-\Im E_2)t}$ and hence
$t_c^{(k)}=\ln|c_2/c_k|/(\Im E_k-\Im E_2)\sim10^3$,
consistent with the observed crossover. For a longer evolving time, the small $\delta_{01}$ keeps the two near-degenerate states dynamically relevant, while their finite Re$E$ splitting generates strong interference oscillations [inset of Fig.~\ref{fig:Quench}(a) and Supplemental Material Sec.~\ref{sec:sm-eigenenergies}]. By contrast, bosons and pseudofermions lack this isolated spectral-edge hierarchy [Figs.~\ref{anyonPT}(a,b)], so more eigenstates affect the short-time dynamics, producing strong fluctuations of $C_{A_LB_1}$ (Supplemental Material, Sec.~\ref{sec:sm-quench}).}

The stability of anyonic dynamics is further apparent in their fidelity evolution, quantified by the normalized overlap function between the initial and final states,
\begin{align}
P(t)=|\langle\psi(t) | \psi_{\rm ini}\rangle|.
\end{align}
Figure ~\ref{fig:Quench}(b) shows
$P(t)\approx 1$ for anyons {across an extended initial period}, reflecting the domination of $|\psi_{\rm ini}\rangle$ in the short-time dynamics; while bosons and pseudofermions show an oscillated dynamics with contributions from multiple states. \blue{The late-time \(P(t)\) is set by initial-state overlaps with the dominant post-quench eigenstates, not by particle statistics.}
Finally, we note that these dynamical behaviors are qualitatively the same for different hopping parameters and anyonic phases, as demonstrated in Supplemental Secs.~\ref{sec:sm_imag_spacing} and \ref{sec:sm-stability}.

\noindent \emph{{\clr Conclusion}.---} 
We uncover criticality induced by anyonic statistics in a non-Hermitian system, {marked by a transition towards a dense spectrum in the Im$E$ axis}. It also highlights the crucial role of {the effective nonlocality in quasi-1D systems, which allows for} a string of {anyonic phases} set by the particle occupations throughout the system.
Although the main text focuses on the two-particle sector for clarity, the robustness of AIST and the associated critical behavior \emph{persist} for more particles, as demonstrated in Supplemental Sec.~\ref{sec:sm-eigenenergies}. Beyond inducing criticality, anyonic statistics is also found to fundamentally reshape the imaginary-spectral structure,
leading to stabilized short-time dynamics in quantum quench.

We emphasize that AIST is fundamentally distinct from the real-complex transition in many-body CSE, induced by the feedback amplification among different NHSE channels, analogous to the single-particle CSE. \blue{The AIST, in contrast, generates complex eigenenergies through the occupation-dependent statistical phase associated with the anyonic Jordan--Wigner string, a mechanism that operates exclusively in the many-body scenario.} The resulting transition and criticality therefore constitute an intrinsic many-body phenomenon with no single-particle counterpart. 
\blue{
We further realize these dynamical signatures on a quantum computer (see End Matter)~\cite{koh2025interacting,zhang2025observation,shen2024enhanced,shen2025observation,XiaoLei2024PRL,WenjinPRA2019,dogra2021quantum,kivela2024quantum}.
They may also be accessible in ultracold atoms, where non-Hermiticity~\cite{liang2022dynamic,zhao2025two} and Abelian anyonic statistics~\cite{Joyce2023arxiv,dhar2025N} have been engineered separately.}

\emph{{\clr Acknowledgements}.---}
L. L. is supported by the National Natural Science Foundation of China (Grant No. 12474159) and the Guangdong Provincial Quantum Science
Strategic Initiative (Grant No. GDZX2504003 and No. GDZX2504006). Y. S. A. is supported by the Kwan Im Thong Hood Cho Temple Early Career Chair Professorship and the Singapore Ministry of Education Academic Research Fund (AcRF) Tier 2 Grant under the award number (MOE-T2EP50224-0021). C.H.L. and Y.Q. acknowledge support from the Singapore Ministry of Education Academic Research Fund Tier 2 under Grant Nos. MOE-T2EP50224-0007, MOE-T2EP50125-0019, and MOE-T2EP50224-0021.

%

\newpage
\clearpage
\section*{End Matter}
\begingroup
\emph{{\clr Critical coupling and average entanglement for anyons}.---}
Here, the criticality of the AIST does not refer to a conventional Hermitian
gap-closing transition characterized by universal correlation or entanglement
scaling. Instead, it refers to the finite-size collapse of the characteristic
inter-sublattice coupling strength associated with collective spectral
complexification. We first distinguish the AIST from the bosonic many-body CSE.  Although both
may exhibit a characteristic scale that decreases with system size, the AIST
originates from occupation-dependent statistical phases that invalidate
pseudo-Hermiticity, rather than from competition between different NHSE
localization channels.  We quantify this distinction using the full-spectrum
scale
\begin{equation}
 \Gamma_{\rm rms}(L,J_p)=\left[D(L)^{-1}\sum_n
 \bigl({\rm Im}\,E_n(L,J_p)\bigr)^2\right]^{1/2},
 \label{eq:gamma-rms-definition}
\end{equation}
where \(D(L)\) is the Hilbert-space dimension at fixed particle number; hence
\(\Gamma_{\rm rms}\) measures the typical imaginary-energy scale of the full
spectrum rather than that of a selected eigenvalue.
Figure~\ref{fig:endmatter-spectral-entanglement} contrasts the sharp,
size-dependent bosonic threshold with the anyonic spectrum, which is already
complex at weak coupling and evolves smoothly with \(J_p\).  Its saturation
scale is also roughly one order of magnitude larger than in the bosonic case,
making the two complexification mechanisms readily distinguishable.
Perturbatively,
\(H_{\rm eff}=H_0+J_pV_{AB}\) has no first-order correction within a fixed
sublattice-particle subspace because \(V_{AB}\) changes the particle numbers on
the two legs.  The first process returning to the original subspace is therefore
second order.  In a biorthogonal eigenbasis of \(H_0\), it has the form
\begin{equation}
 \delta E_n^{(2)}=J_p^2\sum_{m\ne n}
 \frac{\langle\phi_n^L|V_{AB}|\phi_m^R\rangle
       \langle\phi_m^L|V_{AB}|\phi_n^R\rangle}
      {E_n^{(0)}-E_m^{(0)}}.
 \label{eq:end-second-order}
\end{equation}
The occupation-dependent anyonic Peierls phases make this correction
generically complex, hence \(\Gamma_{\rm rms}\propto J_p^2\).  For bosons,
the perturbative coefficients remain real and \(\Gamma_{\rm rms}\) stays zero
until a finite-size spectral threshold is reached (see the Supplemental
Material Sec.~\ref{app:aist-finite-size-diagnostics}).

\begin{figure}[!t]
 \centering
 \includegraphics[width=0.9\linewidth]{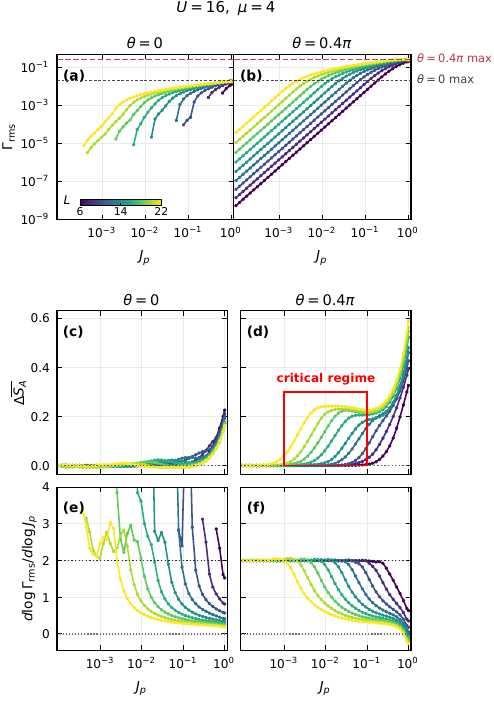}
 \caption{\label{fig:endmatter-spectral-entanglement}
 Spectral and entanglement diagnostics at \(U=16\) and \(\mu=4\).
 (a,b) \(\Gamma_{\rm rms}\): the bosonic spectrum has a sharp threshold,
 whereas the anyonic spectrum complexifies already at weak coupling;
 bosonic data below \(10^{-7}\) are omitted.  (c,d) Entropy response
 \(\Delta\overline S_A\), with the red box marking its fastest anyonic growth.
 (e,f) \(d\log\Gamma_{\rm rms}/d\log J_p\); in the same window, the anyonic
 exponent leaves its weak-coupling value 2.}
\end{figure}

As an independent diagnostic, we bipartition each normalized eigenstate into
the upper and lower legs.  Reshaping its Fock amplitudes into \(C_n\) gives
\(\rho_{A,n}=C_nC_n^\dagger\) and
\(S_A^{(n)}=-{\rm Tr}(\rho_{A,n}\log\rho_{A,n})\).  We use
\begin{equation}
 \Delta\overline S_A=\overline S_A(L,J_p)-\overline S_A(L,10^{-4}),
 \quad
 \chi_S=\partial_{\log J_p}\Delta\overline S_A,
 \label{eq:entropy-response-definition}
\end{equation}
where \(\overline S_A=D^{-1}\sum_nS_A^{(n)}\), and define \(J_S^c(L)\) as
the first clear internal maximum of \(\chi_S\) beyond the weak-coupling
baseline.  Equivalently, it satisfies
\(\partial_{\log J_p}\chi_S=0\) with negative second derivative.  A stable
sequence \(J_S^c(L)\to0\) therefore signals that the fastest entanglement
growth moves to infinitesimal coupling in the thermodynamic limit.  The
anyonic maxima show precisely this monotonic drift, whereas the irregular
bosonic extrema identify only finite-size crossovers
[Fig.~\ref{fig:endmatter-susceptibility-scaling}].  Moreover, the anyonic
entropy grows in the same window in which
\(d\log\Gamma_{\rm rms}/d\log J_p\) leaves its perturbative value 2, showing
that the spectral and entanglement diagnostics locate the same finite-size
AIST regime.  For bosons this exponent is discarded below
\(\Gamma_{\rm rms}=10^{-7}\), where numerical noise dominates.
Numerically, \(\chi_S\) is evaluated on the logarithmic \(J_p\) grid, and the
same first-internal-maximum rule is applied at every \(L\). This common prescription makes the contrast
between the smooth anyonic sequence and the scattered bosonic points a
property of the response rather than of size-dependent fitting choices.

This drift follows from the opposite skin localization of the two legs.
The similarity transformation gives
\(\kappa_\sigma=\frac12\ln|J_R^\sigma/J_L^\sigma|\), so that here
\(J_R^A=J_L^B=1\), \(J_L^A=J_R^B=0.5\), and thus
\(\kappa_A=-\kappa_B=\frac12\ln2\).  An inter-leg transfer at position
\(x_j\), measured from the ladder center, acquires
\(e^{(\kappa_A-\kappa_B)x_j}\).  Since \(|x_j|\lesssim L/2\), its largest
skin-enhanced matrix elements scale as
\(J_pe^{|\kappa_A-\kappa_B|L/2}\); consequently the natural scaling variable
and critical scale are
\begin{equation}
 \begin{aligned}
 g_L&=J_pe^{\kappa_{\rm rel}L},&
 J_S^c(L)&\simeq g_c e^{-\kappa_{\rm rel}L},\\
 \kappa_{\rm rel}&=\frac{|\kappa_A-\kappa_B|}{2}
 =\frac{\ln2}{2}\simeq0.347.
 \end{aligned}
 \label{eq:end-critical-scaling}
\end{equation}
Thus the coefficient \(0.347\) is fixed analytically (see Supplementary material Sec.~\ref{app:aist-finite-size-diagnostics}) by the relative skin
localization rather than obtained from a numerical fit.  It agrees with the
fitted exponents \(0.336\) and \(0.329\) for
\(\theta=0.4\pi\) and \(0.8\pi\), respectively.  This inference requires a
nonzero leading hybridization and a stable susceptibility maximum. At the bosonic and pseudofermionic limits, the nonmonotonic extrema do not support the same single-parameter scaling.

\begin{figure}[!t]
 \centering
 \includegraphics[width=\linewidth]{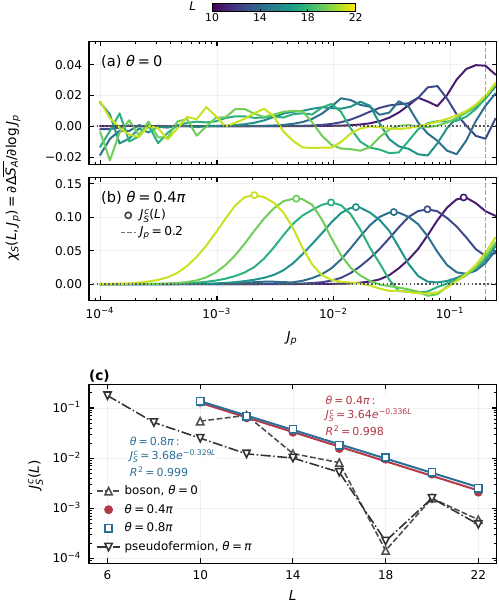}
 \caption{\label{fig:endmatter-susceptibility-scaling}
 Entanglement critical scale.  (a,b) \(\chi_S(L,J_p)\); its first internal
 maximum defines \(J_S^c(L)\).  Anyonic peaks shift systematically to smaller
 \(J_p\), unlike the bosonic extrema; the dashed line marks \(J_p=0.2\).
 (c) The fitted anyonic exponents, \(0.336\) and \(0.329\), agree with the analytically predicted value \(\kappa_{\rm rel}=(\ln 2)/2\simeq 0.347\), while the bosonic ($\theta=0$) and pseudofermionic (\(\theta=\pi\)) points show no monotonic scaling.}
\end{figure}

\emph{{\clr Quantum-hardware realization}.---}
To test whether the predicted short-time stabilization is accessible on present-day quantum processors, we implement the representative post-quench dynamics of the $L=10$, $N=2$ ladder on the IBM superconducting processor \texttt{ibm\_berlin}. Restriction to the \(D=210\) two-particle case
compresses the 20-mode ladder to eight qubits; the final state reaches about \(0.99\) fidelity for the anyonic states.  Working directly in
this symmetry-resolved sector permits shallow state preparation and avoids
ancilla qubits.  The survival overlap
\(P(t)=|\langle\psi_{\rm ini}|\psi(t)\rangle|\) is read from the all-zero
probability after applying the inverse reference-state circuit.  We use 11
times up to \(t=5000\), \(2\times10^4\) shots per circuit, M3 readout
correction, and zero-noise extrapolation; circuit and mitigation details are
given in the Supplemental Material Sec.~\ref{app:qc-details}.  The same encoding, circuit family, and
readout are used for the anyonic trajectory and the bosonic reference, making
the latter a direct control for the role of exchange statistics.

\begin{figure}[!t]
 \centering
 \includegraphics[width=0.78\linewidth]{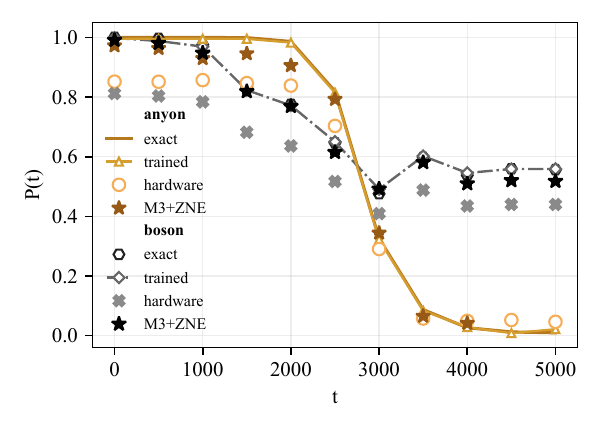}
 \caption{\label{fig:qc-hardware-demo}
 Survival overlap on \texttt{ibm\_berlin}, using the same eight-qubit encoding
 and readout for \(\theta=0.4\pi\) and the bosonic control.  After M3 readout
 correction and zero-noise extrapolation, the anyonic trajectory retains the
 predicted short-time high-overlap window before decaying, unlike the bosonic
 evolution. All other parameters are the same as in Fig.~4.}
\end{figure}

As Fig.~\ref{fig:qc-hardware-demo} shows, the hardware data clearly resolve the
statistics-dependent dynamical contrast: the anyonic state retains the predicted
short-time high-survival window before decaying, whereas the bosonic control
shows a qualitatively different response. Beyond the present model, the same
compact Fock-space encoding, and overlap-readout framework
can be adapted to other few-particle lattice models and quench protocols by
changing only the target states and circuit parameters, providing a transferable
route to related interacting and non-Hermitian dynamics on programmable quantum
processors.
\endgroup

\clearpage
\onecolumngrid
\begin{center}
\textbf{\large Supplementary Materials}
\end{center}

\setcounter{section}{0}
\setcounter{subsection}{0}
\setcounter{equation}{0}
\setcounter{figure}{0}
\setcounter{table}{0}
\renewcommand{\thesection}{S\arabic{section}}
\renewcommand{\theequation}{S\arabic{equation}}
\renewcommand{\thefigure}{S\arabic{figure}}
\renewcommand{\thetable}{S\arabic{table}}
\renewcommand{\theHsection}{supp.\arabic{section}}
\renewcommand{\theHequation}{supp.\arabic{equation}}
\renewcommand{\theHfigure}{supp.\arabic{figure}}
\renewcommand{\theHtable}{supp.\arabic{table}}
\renewcommand{\bibnumfmt}[1]{[S#1]}
\renewcommand{\citenumfont}[1]{S#1}

\section{Anyonic Jordan--Wigner Mapping}
\label{sec:sm-jw}

For completeness, we briefly review the anyonic Jordan--Wigner mapping~\cite{SLiu2018PRL, SLange2017PRL, Sqin2025CP, SBonk2025PRL} which establishes
the isomorphism between 1D anyons and non-local bosons. In particular, we show that the operators \(\hat a_i\),
defined via the nonlocal mapping
\begin{equation}
\hat a_j = \hat b_j \exp\Big(-i\theta \sum_{k<j} \hat n_k\Big),
\end{equation}
indeed satisfy the anyonic commutation relations
\begin{equation}
\hat a_i \hat a_j^\dagger - e^{-i\theta\,\mathrm{sgn}(i-j)} \hat a_j^\dagger \hat a_i = \delta_{ij},
\end{equation}
provided that the underlying particles \(\hat b_i\) obey bosonic commutation relations.

We then give a short proof. For \(i<j\), we rewrite the product of anyonic operators in terms of the bosonic ones:
\begin{align}
\hat a_i \hat a_j^{\dagger}
&=
\hat b_i
e^{-i\theta \sum_{k<i}\hat n_k}
e^{+i\theta \sum_{k<j}\hat n_k}
\hat b_j^{\dagger}
\nonumber\\
&=
\hat b_i
e^{+i\theta\left(\hat n_i+\sum_{i<k<j}\hat n_k\right)}
\hat b_j^{\dagger}
\nonumber\\
&=
e^{+i\theta\sum_{i<k<j}\hat n_k}
\hat b_i e^{+i\theta\hat n_i}\hat b_j^{\dagger}
\nonumber\\
&=
e^{+i\theta}
e^{+i\theta\sum_{i<k<j}\hat n_k}
e^{+i\theta\hat n_i}
\hat b_i\hat b_j^{\dagger}
\nonumber\\
&=
e^{+i\theta}
e^{+i\theta\sum_{i<k<j}\hat n_k}
e^{+i\theta\hat n_i}
\hat b_j^{\dagger}\hat b_i ,
\\
e^{-i\theta\,\mathrm{sgn}(i-j)} \hat a_j^{\dagger} \hat a_i
&=
e^{+i\theta}
e^{+i\theta \sum_{k<j}\hat n_k}
\hat b_j^{\dagger}\hat b_i
e^{-i\theta \sum_{k<i}\hat n_k}
\nonumber\\
&=
e^{+i\theta}
e^{+i\theta\sum_{i<k<j}\hat n_k}
e^{+i\theta\hat n_i}
\hat b_j^{\dagger}\hat b_i .
\end{align}

Subtracting the two expressions yields
\begin{align}
\hat a_i \hat a_j^{\dagger}
-
e^{-i\theta\,\mathrm{sgn}(i-j)}
\hat a_j^{\dagger}\hat a_i
&=
e^{+i\theta}
e^{+i\theta\sum_{i<k<j}\hat n_k}
e^{+i\theta\hat n_i}
\left(
\hat b_j^{\dagger}\hat b_i
-
\hat b_j^{\dagger}\hat b_i
\right)
\nonumber\\
&=0,
\end{align}
which verifies the anyonic commutation relations for \(i<j\).

The proof for \(i>j\) follows analogously. For \(i=j\), one simply notes that
\(\hat a_i^{\dagger} \hat a_i = \hat b_i^{\dagger} \hat b_i\), \(\hat a_i \hat a_i^{\dagger}=\hat b_i\hat b_i^\dagger\), and \(e^{-i\theta\,\mathrm{sgn}(0)}=1\). Therefore,
\begin{equation}
\hat a_i \hat a_i^\dagger-\hat a_i^\dagger\hat a_i
=
\hat b_i\hat b_i^\dagger-\hat b_i^\dagger\hat b_i
=
1 .
\end{equation}
Therefore, the mapping correctly reproduces the anyonic algebra from bosonic operators.

\begingroup
\color{black}

\paragraph*{Boundary conditions under the anyonic Jordan--Wigner mapping.}
The Jordan--Wigner mapping requires additional care under periodic boundary
conditions. To illustrate this point, consider a chain of \(L\) sites with
periodic boundary conditions imposed in the anyonic representation. For a bulk
link, \(1\leq j<L\), the hopping operator transforms as
\begin{align}
\hat a_j^\dagger \hat a_{j+1}
&=
e^{+i\theta\sum_{k<j}\hat n_k}
\hat b_j^\dagger \hat b_{j+1}
e^{-i\theta\sum_{k<j+1}\hat n_k}
\nonumber\\
&=
\hat b_j^\dagger \hat b_{j+1}
e^{-i\theta\hat n_j}.
\end{align}
Thus, the anyonic statistical phase appears as an occupation-dependent phase
dressing the bosonic hopping.

The boundary link connecting sites \(L\) and \(1\), however, transforms
differently. Defining the conserved total particle number as
\begin{equation}
\hat N=\sum_{j=1}^{L}\hat n_j,
\end{equation}
we obtain
\begin{align}
\hat a_L^\dagger \hat a_1
&=
e^{+i\theta\sum_{k<L}\hat n_k}
\hat b_L^\dagger \hat b_1
\nonumber\\
&=
e^{+i\theta(\hat N-\hat n_L)}
\hat b_L^\dagger \hat b_1
\nonumber\\
&=
e^{+i\theta(\hat N-1)}
\hat b_L^\dagger \hat b_1
e^{-i\theta\hat n_L}.
\end{align}
Similarly, the hopping in the opposite direction becomes
\begin{equation}
\hat a_1^\dagger \hat a_L
=
e^{-i\theta(\hat N-1)}
e^{+i\theta\hat n_L}
\hat b_1^\dagger \hat b_L.
\end{equation}

Therefore, periodic boundary conditions in the anyonic representation do not
correspond to naive periodic boundary conditions,
\(\hat b_{L+1}=\hat b_1\), in the bosonic representation. Within a fixed-\(N\)
particle-number sector, the boundary term can equivalently be represented by
the twisted boundary condition
\begin{equation}
\hat b_{L+1}
=
e^{+i\theta(N-1)}\hat b_1,
\end{equation}
for the Jordan--Wigner convention adopted here. The hopping in the opposite
direction acquires the complex-conjugate twist. Consequently, the spectra
obtained by independently imposing periodic boundary conditions in the
anyonic and bosonic representations need not coincide unless the corresponding
particle-number-dependent boundary twist is included.

Under open boundary conditions, the link between sites \(L\) and \(1\) is
absent, and no Jordan--Wigner string crosses the boundary. The boundary twist
therefore does not arise. Since the present work focuses exclusively on open
boundary conditions, the mapping can be applied directly without introducing
an additional boundary term.

\endgroup

\section{Perturbation theory applied to different energy clusters}
\label{sec:sm-perturbation}

To explain the two-stage complex plateau observed in Fig.~\ref{N0ReEvsJp} of the main text, we present a perturbative analysis in this section. The calculation indicates that the difference arises from the leading interchain-coupling-induced, statistics-sensitive processes in different subspaces, namely, a fourth-order term of $\mathcal O(J^2J_p^2)$ for two particles occupying the same chain, and a second-order term of $\mathcal O(J_p^2)$ for them occupying different chains, as elaborated in the following subsections.

For a self-contained discussion, we begin with the anyonic Hamiltonian, 
\begin{align}
\hat{H}= & -J\sum_{j=1}^{L-1}\sum_{\sigma=A,B} \left(e^{\alpha_\sigma} \hat{a}_{j, \sigma}^{\dagger} \hat{a}_{j+1, \sigma}+e^{-\alpha_\sigma} \hat{a}_{j+1, \sigma}^{\dagger} \hat{a}_{j, \sigma}\right) \notag \\
& +J_p \sum_{j=1}^L\left(\hat{a}_{j, A}^{\dagger} \hat{a}_{j, B}+\hat{a}_{j, B}^{\dagger} \hat{a}_{j, A}\right)
+\mu \sum_{j=1}^L\left(\hat{n}_{j, A}-\hat{n}_{j, B}\right)\notag \\
& +\frac{U}{2} \sum_{j=1}^L\sum_{\sigma=A,B} \hat{n}_{j, \sigma}\left(\hat{n}_{j, \sigma}-1\right).
\label{eq:H}
\end{align}
The anyonic creation/annihilation operators satisfy \begin{equation}
\begin{array}{l}
{[{{\hat a}_j},{{\hat a}_k}]_\theta } \equiv {{\hat a}_j}{{\hat a}_k} - {e^{ i\theta {\mathop{\rm sgn}} (j - k)}}{{\hat a}_k}{{\hat a}_j} = 0,\\
{[{{\hat a}_j},\hat a_k^\dag ]_{ - \theta }} \equiv {{\hat a}_j}\hat a_k^\dag  - {e^{-i\theta {\mathop{\rm sgn}} (j - k)}}\hat a_k^\dag {{\hat a}_j} = {\delta _{jk}},\label{sm:commute}
\end{array}
\end{equation}.

We rewrite the Hamiltonian as $\hat{H}=\hat{H}_0+\hat{V}$,
\begin{align}
\hat H_0=\hat H_\mu+\hat H_U, \qquad
\hat V=\hat H_J+\hat H_{J_p},
\label{HoHv}
\end{align}
with $\hat H_\mu=\mu\sum_j(\hat n_{j,A}-\hat n_{j,B})$ and $\hat H_U=\frac{U}{2}\sum_{j,\sigma}\hat n_{j,\sigma}(\hat n_{j,\sigma}-1)$. $\hat H_J$ and $\hat H_{J_p}$ denote the intra-chain and inter-chain hopping terms, respectively.

In the two-particle sector, a convenient (unnormalized) basis is
\begin{align}
|j,\sigma; k,\sigma'\rangle \equiv \hat a^\dagger_{j,\sigma}\hat a^\dagger_{k,\sigma'}|0\rangle,
\end{align}
with $j=k$ and $\sigma'=\sigma$ for on-site doublons.

\subsection{ Perturbation theory in the lowest $BB$ scattering subspace (two particles)}

The key result of this subsection is that the leading $J_p$-dependent, statistics-sensitive plaquette-assisted exchange enters at fourth order, as captured by Eq.~\eqref{eq:path_exchange}. 

For scattering states ($j\neq k$) the interaction part does not contribute, hence
\begin{align}
\hat H_0|j,\sigma;k,\sigma'\rangle
=E^{(0)}_{\sigma\sigma'}|j,\sigma;k,\sigma'\rangle,
\qquad
E^{(0)}_{\sigma\sigma'}=
\big(s_\sigma+s_{\sigma'}\big)\mu,
\quad
s_A=+1,\ s_B=-1.
\end{align}
Thus
\begin{align}
E^{(0)}_{AA}=+2\mu,\qquad
E^{(0)}_{AB}=0,\qquad
E^{(0)}_{BB}=-2\mu.
\end{align}
Assuming $\mu>0$, the lowest scattering manifold is the $BB$ sector. Define the projector onto the \emph{lowest $BB$ scattering subspace}
\begin{align}
P\equiv \sum_{1 \leq j<k \leq L} |j,B;k,B\rangle\langle j,B;k,B|,
\qquad Q\equiv 1-P.
\end{align}
Here we keep only one ordering, $j<k$, because
$|k,B;j,B\rangle=e^{+i\theta}|j,B;k,B\rangle$ for $k>j$ and therefore the two orderings are not independent states. We denote the unperturbed energy in this manifold by $E_0=-2\mu$.

The interchain term $\hat H_{J_p}$ flips the chain index on a single site:
\begin{align}
\hat H_{J_p}:\quad B\leftrightarrow A.
\end{align}
Therefore, any process that starts in the $BB$ subspace and ends in the $BB$ subspace must contain an \emph{even} number of $\hat H_{J_p}$ insertions. In particular, all contributions containing an odd power of $J_p$ vanish after projection back to $P$:
\begin{align}
P(\cdots \hat H_{J_p}\cdots)P=0 \quad \text{if the total number of }\hat H_{J_p}\text{ is odd.}
\end{align}
This is a selection rule on the number of interchain hoppings, not on the total perturbative order; odd-order terms containing an even number of $\hat H_{J_p}$ can still occur.

Using (degenerate) Schrieffer--Wolff / Brillouin--Wigner expansion~\cite{SPhysRevC99,SFieldsGuptaVary1996} 
, the projected effective Hamiltonian reads
\begin{align}
\hat H_{\mathrm{eff}}
= P\hat H_0P + P\hat V P
- P\hat V Q\frac{1}{\hat H_0-E_0}Q\hat V P
+ P\hat V Q\frac{1}{\hat H_0-E_0}Q\hat V Q\frac{1}{\hat H_0-E_0}Q\hat V P
+\cdots
\end{align}
Within $P$, $P\hat H_0P=E_0 P$ is a constant shift.

Since $\hat H_{J_p}$ exchanges the chain indexes, $P\hat H_{J_p}P=0$. Hence
\begin{align}
\hat H_{\mathrm{eff}}^{(1)}=P\hat H_J P,
\end{align}
which generates the kinetic motion of two particles on the $B$ chain (with non-reciprocal hopping $e^{\pm\alpha_B}$). No exchange phase $\theta$ appears at this order. At third order, terms with an odd number of $H_{J_p}$ vanish by the chain-parity selection rule. Terms of order $J J_p^2$ are allowed, but contain only one spatial hop; the statistical phases from the two interchain vertices cancel, so they only renormalize the $BB$ kinetic motion and are $\theta$ independent. The irreducible $J^3$ contribution vanishes because a $BB$ doublon created after leaving $P$ returns to the $BB$ scattering manifold after the next spatial hop. Consequently, the leading $J_p$-dependent, $\theta$-dependent plaquette exchange process requires two $H_{J_p}$ insertions and two spatial hoppings, i.e., it first arises at $\mathcal{O}(J^2J_p^2)$. 

At second order, only the two-channel loops $J^2$ and $J_p^2$ survive:
\begin{align}
\hat H_{\mathrm{eff}}^{(2)}=
- P\hat H_J Q\frac{1}{\hat H_0-E_0}Q\hat H_J P
- P\hat H_{J_p} Q\frac{1}{\hat H_0-E_0}Q\hat H_{J_p} P.
\end{align}
\begin{itemize}
\item The $J^2$ term proceeds via a $BB$ doublon intermediate state, e.g. $|j,B;j+1,B\rangle\to |j,B;j,B\rangle \to |j,B;j+1,B\rangle$, with energy denominator
\begin{align}
\Delta_U=\big(U-2\mu\big)-(-2\mu)=U.
\end{align}
It generates intrachain virtual corrections within the $BB$ sector, which are already present at $J_p=0$ and are therefore not related to the interchain-coupling-induced mechanism discussed here.
\item The $J_p^2$ term proceeds via an $AB$ scattering intermediate state, $|BB\rangle\to |AB\rangle\to |BB\rangle$, with denominator
\begin{align}
\Delta_\mu=0-(-2\mu)=2\mu,
\end{align}
\end{itemize}
leading to a uniform shift within the $BB$ scattering manifold (no exchange structure, no $\theta$). Therefore, the second-order $J_p$-dependent contribution within $P$ is independent of $\theta$.

We next provide an explicit fourth-order virtual process that generates a $J_p$-dependent anyonic exchange phase without ever visiting an intrachain doublon (hence without any $U$-denominators). Because the two orderings of identical $B$-chain anyons are not independent, the exchange must be represented as a closed virtual loop that returns to the same physical $BB$ configuration. We therefore take the canonical nearest-neighbor state
$
|i\rangle \equiv |j,B;\,j\!+\!1,B\rangle
$
with $j<j+1$. The irreducible fourth-order $Q$-space contribution contains
\begin{equation}
\hat H_{\mathrm{eff}}^{(4)} \supset
P\,\hat V Q\frac{1}{E_0-\hat H_0}Q\hat V Q\frac{1}{E_0-\hat H_0}Q\hat V Q\frac{1}{E_0-\hat H_0}Q\hat V\,P ,
\end{equation}
with $E_0=-2\mu$ and $\hat V=\hat H_J+\hat H_{J_p}$.

Starting from $|i\rangle$, one minimal ``plaquette'' sequence on the ladder is
\[
(j,B;\,j\!+\!1,B)
\ \xrightarrow{\,J_p \text{ at } j\,}\
(j,A;\,j\!+\!1,B)
\ \xrightarrow{\,J \text{ on } B\,}\
(j,A;\,j,B)
\ \xrightarrow{\,J \text{ on } A\,}\
(j\!+\!1,A;\,j,B)
\ \xrightarrow{\,J_p \text{ at } j\!+\!1\,}\
(j,B;\,j\!+\!1,B)=|i\rangle .
\]
Denoting the intermediate states by
$
|m_1\rangle\equiv |j,A;\,j\!+\!1,B\rangle,
$
$
|m_2\rangle\equiv \hat a^\dagger_{j,A}\hat a^\dagger_{j,B}|0\rangle,
$
and
$
|m_3\rangle\equiv |j\!+\!1,A;\,j,B\rangle,
$
their unperturbed energies under $\hat H_0=\hat H_\mu+\hat H_U$ are all $E_{m_1}=E_{m_2}=E_{m_3}=0$ (they belong to the $AB$ sector and carry no intrachain doublon penalty), while $E_0=-2\mu$ in the $BB$ manifold. Hence each energy denominator equals
\begin{equation}
E_0-E_{m_\ell}=-2\mu,\qquad \ell=1,2,3.
\end{equation}
Using the global ordering $A:1,\ldots,L$ followed by $B:L+1,\ldots,2L$, the corresponding vertex matrix elements are
\begin{equation}
\langle m_1|\hat H_{J_p}|i\rangle = J_p,
\qquad
\langle m_2|\hat H_J|m_1\rangle = -J e^{\alpha_B},
\end{equation}
\begin{equation}
\langle m_3|\hat H_J|m_2\rangle = -J e^{-\alpha_A},
\qquad
\langle i|\hat H_{J_p}|m_3\rangle = J_p e^{+i\theta}.
\end{equation}
The anyonic phase is accumulated by the closed exchange loop; in this canonical ordering it appears at the final interchain hop. Collecting all factors, this virtual path contributes
\begin{equation}
\langle i|\hat H_{\mathrm{eff}}^{(4)}|i\rangle
\supset
-\frac{J^2J_p^2}{8\mu^3}\,e^{(\alpha_B-\alpha_A)}\,e^{+i\theta}.
\label{eq:path_exchange}
\end{equation}
The oppositely oriented plaquette sequence contributes
$-J^2J_p^2 e^{(\alpha_A-\alpha_B)}e^{-i\theta}/(8\mu^3)$. Thus the pair of closed exchange loops contains the combination
\begin{equation}
-\frac{J^2J_p^2}{8\mu^3}
\left[
 e^{(\alpha_B-\alpha_A)}e^{+i\theta}
 +e^{(\alpha_A-\alpha_B)}e^{-i\theta}
\right].
\end{equation}
For $\alpha_A=\alpha$ and $\alpha_B=-\alpha$, this contribution is generically complex when $\alpha\neq0$ and $\theta\neq0,\pi$. The important point for the present argument is its perturbative order, $\mathcal O(J^2J_p^2)$; additional fourth-order paths can modify the numerical prefactor but not this order counting.

The expansion is controlled when the virtual gaps associated with leaving $P$ are large compared with the perturbations:
\begin{align}
U\gg J,\qquad 2\mu\gg J,J_p,
\end{align}
and $\mu>0$ to ensure the $BB$ scattering manifold is the lowest-energy scattering sector.

In summary, with $H_0=H_\mu+H_U$ and projection to the lowest $BB$ scattering subspace, (i) the second-order $J_p$-dependent contribution is independent of $\theta$; (ii) the allowed third-order $J J_p^2$ processes are also $\theta$ independent; and (iii) the leading $J_p$-dependent, $\theta$-dependent interchain plaquette exchange contribution arises at fourth order, $\mathcal{O}(J^{2}J_{p}^{2})$. The two orderings $\lvert j,B;k,B\rangle$ and $\lvert k,B;j,B\rangle$ represent the same physical state up to the anyonic exchange phase, so the statistical effect is correctly encoded by closed virtual exchange loops rather than by an off-diagonal coupling between these two orderings.

\subsection{Second-order perturbation theory in the $AB$ two-particle sector}

The key result of this subsection is that the earliest effective signature of anyonic statistics already appears in the \(AB\) two-particle sector through the second-order exchange amplitude in Eq.~\eqref{EABsec}.

We focus on the part of the two-particle $AB$ sector with one particle on each chain and on different rungs,
\begin{equation}
\mathcal H_{AB}
=
\mathrm{span}\big\{
|j,k\rangle_{AB}
=
\hat a^\dagger_{j,A}\hat a^\dagger_{k,B}|0\rangle
\;\big|\;
j\neq k
\big\}.
\end{equation}
Throughout, we adopt the real-space embedding
\begin{equation}
A:\;1,\dots,L,
\qquad
B:\;L+1,\dots,2L,
\end{equation}
so that any reordering between $A$- and $B$-chain operators carries an anyonic phase.

The unperturbed Hamiltonian is chosen as $\hat H_0=\hat H_\mu+\hat H_U$ as defined in eq.~\ref{HoHv}.
For states $|j,k\rangle_{AB}$ with $j\neq k$, there is no on-site interaction, and the chemical potential contributions cancel,
\begin{equation}
E^{(0)}_{AB}=0.
\end{equation}
The relevant intermediate states generated by inter-chain tunneling are
\begin{align}
E^{(0)}_{AA} &= +2\mu, \\
E^{(0)}_{BB} &= -2\mu .
\end{align}

We focus on the inter-chain hopping
\begin{equation}
V \equiv H_{J_p}
=
J_p\sum_{\ell}
\left(
\hat a^\dagger_{\ell,A}\hat a_{\ell,B}
+
\hat a^\dagger_{\ell,B}\hat a_{\ell,A}
\right)
\end{equation}
and treat it as a perturbation (the intra-chain hopping $H_J$ can be omitted since in the $AB$ two-particle scattering sector, it alone does not produce a net $\theta$-dependence because it cannot generate an exchange process that changes the particle ordering.)
The second-order effective Hamiltonian within $\mathcal H_{AB}$ is given by
\begin{equation}
\label{eq:Heff2}
\langle f|H^{(2)}_{\mathrm{eff}}|i\rangle
=
\sum_{m\notin \mathcal H_{AB}}
\frac{
\langle f|V|m\rangle
\langle m|V|i\rangle
}{
E^{(0)}_{AB}-E^{(0)}_m
}.
\end{equation}

We focus on the exchange matrix element between
\begin{equation}
|i\rangle = |j,k\rangle_{AB},
\qquad
|f\rangle = |k,j\rangle_{AB},
\qquad j\neq k.
\end{equation}
There are two virtual processes contributing at second order:
\begin{enumerate}
\item a path via the $BB$ intermediate state,
\item a path via the $AA$ intermediate state.
\end{enumerate}

Because of the anyonic commutation relations and the real-space embedding, operator reordering between $A$ and $B$ chains generates statistical phase factors. A direct evaluation shows that the $BB$ path contributes
\begin{equation}
\langle k,j|H^{(2)}_{\mathrm{eff}}|j,k\rangle_{BB}
=
\frac{J_p^2}{2\mu}e^{+i\theta\,\mathrm{sgn}(j-k)},
\end{equation}
while the $AA$ path contributes
\begin{equation}
\langle k,j|H^{(2)}_{\mathrm{eff}}|j,k\rangle_{AA}
=
-\frac{J_p^2}{2\mu}\,
e^{-i\theta\,\mathrm{sgn}(j-k)}.
\end{equation}

Summing the two contributions, we obtain the total second-order exchange matrix element
\begin{equation}
\label{EABsec}
\langle k,j|H^{(2)}_{\mathrm{eff}}|j,k\rangle
=
\frac{J_p^2}{2\mu}
\left[
e^{+i\theta\,\mathrm{sgn}(j-k)}
-
e^{-i\theta\,\mathrm{sgn}(j-k)}
\right].
\end{equation}

Several important conclusions follow immediately:
\begin{itemize}
\item For $\theta=0$, the contributions from the $AA$ and $BB$ virtual paths cancel exactly, and the second-order exchange matrix element vanishes.
\item For $\theta\neq0,\pi$ and $j\neq k$, a finite exchange coupling is generated already at second order.
\item The statistical phase originates from operator reordering between $A$- and $B$-chain particles under the real-space embedding, rather than from spatial hopping.
\end{itemize}

Thus, when the two particles initially occupy different sites on different chains, inter-chain tunneling induces an effective anyonic exchange at second order, which fundamentally distinguishes the cases $\theta=0,\pi$ and $\theta\neq0,\pi$. Including intra-chain hopping introduces a first-order contribution to the non-Hermitian hopping. Together with the statistical-angle effect, this makes the energy spectrum complex.

\section{Pseudo-Hermiticity Breaking in the Anyonic Ladder Model}
\label{sec:sm-pseudohermiticity}

We consider the two-chain bosonic Hamiltonian derived from the anyonic model using the above transformation, 
\begin{equation}
\begin{aligned}
\hat{H} = & - \sum_{j=1}^{L-1} \sum_{\sigma=A,B} \left( J_{L,\sigma} \hat{b}_{j,\sigma}^\dag e^{-i \theta \hat{n}_{j,\sigma}} \hat{b}_{j+1,\sigma} + J_{R,\sigma} \hat{b}_{j+1,\sigma}^\dag e^{i \theta \hat{n}_{j,\sigma}} \hat{b}_{j,\sigma} \right) \\
& + \sum_{j=1}^L \left( J_p e^{i \theta \Phi_j} \hat{b}_{j,B}^\dag \hat{b}_{j,A} + J_p \hat{b}_{j,A}^\dag \hat{b}_{j,B} e^{-i \theta \Phi_j} \right) + 
\frac{U}{2}\sum_{j=1,\sigma}^L
\hat n_{j,\sigma}\left(\hat n_{j,\sigma}-1\right)
+\mu\sum_{j=1}^L\left(
\hat n_{j,A}-\hat n_{j,B}
\right),
\end{aligned}
\end{equation}
where
\[
\Phi_j = \sum_{k=j}^L \hat{n}_{k,A} + \sum_{k=1}^{j-1} \hat{n}_{k,B},
\quad
J_{L,\sigma} = J e^{\alpha_\sigma},\quad J_{R,\sigma} = J e^{-\alpha_\sigma},\quad \alpha_A = \alpha,\ \alpha_B = -\alpha.
\]

We now define the pseudo-Hermiticity operator as:
\[
\mathcal{K} := \mathcal{R}_z \mathcal{I}  \mathcal{T},
\]
where \(\mathcal{T}\) is time-reversal (complex conjugation), \(\mathcal{I}\) is the spatial inversion operator: \(j \rightarrow L+1-j\), acts identically on both chains, \(\mathcal{R}_z = \exp\left(-i \theta \sum_{j=1}^L \sum_{\sigma=A,B} \frac{\hat{n}_{j,\sigma} (\hat{n}_{j,\sigma} - 1)}{2} \right)\) is a local density-dependent rotation. Action of \(\mathcal{K}\) on field operators:
\[
\mathcal{I} \hat{b}_{j,\sigma} \mathcal{I}^\dag = \hat{b}_{L+1-j,\sigma},\quad
\mathcal{R}_z \hat{b}_{j,\sigma} \mathcal{R}_z^\dag = e^{i \theta \hat{n}_{j,\sigma}} \hat{b}_{j,\sigma},\quad
\mathcal{R}_z \hat{b}_{j,\sigma}^\dag \mathcal{R}_z^\dag = \hat{b}_{j,\sigma}^\dag e^{-i \theta \hat{n}_{j,\sigma}}.
\]

Thus, for the total \(\mathcal{K}\), we obtain:
\[
\mathcal{K} \hat{b}_{j,\sigma} \mathcal{K}^\dag = e^{i \theta \hat{n}_{L+1-j,\sigma}} \hat{b}_{L+1-j,\sigma},\quad
\mathcal{K} \hat{b}_{j,\sigma}^\dag \mathcal{K}^\dag = \hat{b}_{L+1-j,\sigma}^\dag e^{-i \theta \hat{n}_{L+1-j,\sigma}}.
\]

Each chain’s Hamiltonian
\[
\hat{H}_\sigma = - \sum_{j=1}^{L-1} \left( J_{L,\sigma} \hat{b}_{j,\sigma}^\dag e^{-i \theta \hat{n}_{j,\sigma}} \hat{b}_{j+1,\sigma} + J_{R,\sigma} \hat{b}_{j+1,\sigma}^\dag e^{i \theta \hat{n}_{j,\sigma}} \hat{b}_{j,\sigma} \right) + \frac{U}{2} \sum_j \hat{n}_{j,\sigma} (\hat{n}_{j,\sigma} - 1)
\]
satisfies:
\[
\mathcal{K} \hat{H}_\sigma \mathcal{K}^\dag = \hat{H}_\sigma^\dag,
\]
i.e., each single chain is pseudo-Hermitian under \(\mathcal{K}\).

For inter-chain coupling term, we examine:
\[
\hat{H}_{AB} = J_p \sum_{j=1}^L \left( e^{i \theta \Phi_j} \hat{b}_{j,B}^\dag \hat{b}_{j,A} + \text{h.c.} \right),
\quad
\Phi_j = \sum_{k=j}^L \hat{n}_{k,A} + \sum_{k=1}^{j-1} \hat{n}_{k,B}.
\]

Under \(\mathcal{K} = \mathcal{R}_z \mathcal{I} \mathcal{T}\), one finds:
\[
\mathcal{K} \left( \hat{b}_{j,B}^\dag \hat{b}_{j,A} \right) \mathcal{K}^\dag = \hat{b}_{L+1-j,B}^\dag e^{-i \theta \hat{n}_{L+1-j,B}} \cdot e^{i \theta \hat{n}_{L+1-j,A}} \hat{b}_{L+1-j,A},
\]
which gives:
\[
\mathcal{K} \left( \hat{b}_{j,B}^\dag \hat{b}_{j,A} \right) \mathcal{K}^\dag = \hat{b}_{L+1-j,B}^\dag \hat{b}_{L+1-j,A} \cdot e^{i \theta (\hat{n}_{L+1-j,A} - \hat{n}_{L+1-j,B})}.
\]

However, the Peierls phase transforms as:
\[
\Phi_j \rightarrow \Phi_j' = \sum_{k=L+1-j}^L \hat{n}_{k,A} + \sum_{k=1}^{L-j} \hat{n}_{k,B}.
\]
 Hence:
\[
\mathcal{K} \hat{H}_{AB} \mathcal{K}^\dag \neq \hat{H}_{AB}^\dag.
\]

From the above calculations, we note that although each chain individually respects pseudo-Hermiticity under \(\mathcal{K} = \mathcal{R}_z \mathcal{I}\mathcal{T}\), the full two-chain Hamiltonian fails to satisfy it,
\[
\mathcal{K} \hat{H} \mathcal{K}^\dag \neq \hat{H}^\dag,
\]
because the nonlocal Peierls phase \(\Phi_j\) breaks inversion symmetry $\mathcal{I}$ while keeping the others. 

Explicitly, the inter-chain coupling in the double-chain bosonic Hamiltonian reads
\[
\hat{H}_{AB} = J_p \sum_{j=1}^L \left( e^{i \theta \Phi_j} \hat{b}_{j,B}^\dagger \hat{b}_{j,A} + \text{h.c.} \right),
\quad
\Phi_j = \sum_{k=j}^L \hat{n}_{k,A} + \sum_{k=1}^{j-1} \hat{n}_{k,B},
\]
and the symmetry operation yields
\begin{align*}
\mathcal{K} \left( e^{i \theta \Phi_j} \hat{b}_{j,B}^\dagger \hat{b}_{j,A} \right) \mathcal{K}^\dagger
&= \mathcal{R}_z \cdot \mathcal{I} \cdot \mathcal{T} \left( e^{i \theta \Phi_j} \hat{b}_{j,B}^\dagger \hat{b}_{j,A} \right) \mathcal{T}^{-1} \cdot \mathcal{I}^{-1} \cdot \mathcal{R}_z^\dagger \\
&= e^{-i \theta \Phi_j'} \cdot \mathcal{R}_z \left( \hat{b}_{L+1-j,B}^\dagger \hat{b}_{L+1-j,A} \right) \mathcal{R}_z^\dagger \\
&= e^{-i \theta \Phi_j'} \cdot \hat{b}_{L+1-j,B}^\dagger  e^{i \theta (\hat{n}_{L+1-j,A} - \hat{n}_{L+1-j,B})} \hat{b}_{L+1-j,A} ,
\end{align*}
with
\[
\Phi_j' = \sum_{k=L+1-j}^{L} \hat{n}_{k,A} + \sum_{k=1}^{L-j} \hat{n}_{k,B}.
\]
In contrast to the original term, the transformed expression includes an additional operator-valued phase factor $e^{i \theta (\hat{n}_A - \hat{n}_B)}$ and the number-dependent $\Phi_j'$ does not map back to any $\Phi_{j}$ in the original form. As a result,
\[
\mathcal{K} \hat{H}_{AB} \mathcal{K}^\dagger \neq \hat{H}_{AB}^\dagger.
\]
Thus, inter-chain coupling invalidates pseudo-Hermiticity of the total system for anyons, leading to complex eigenenergies not necessarily forming conjugate pairs. This is the novelty of our model in the main text.

\section{Evolution of eigenenergies with interchain coupling}
\label{sec:sm-eigenenergies}

\begin{figure}[htbp]
\centering
\includegraphics[height=6.8cm]{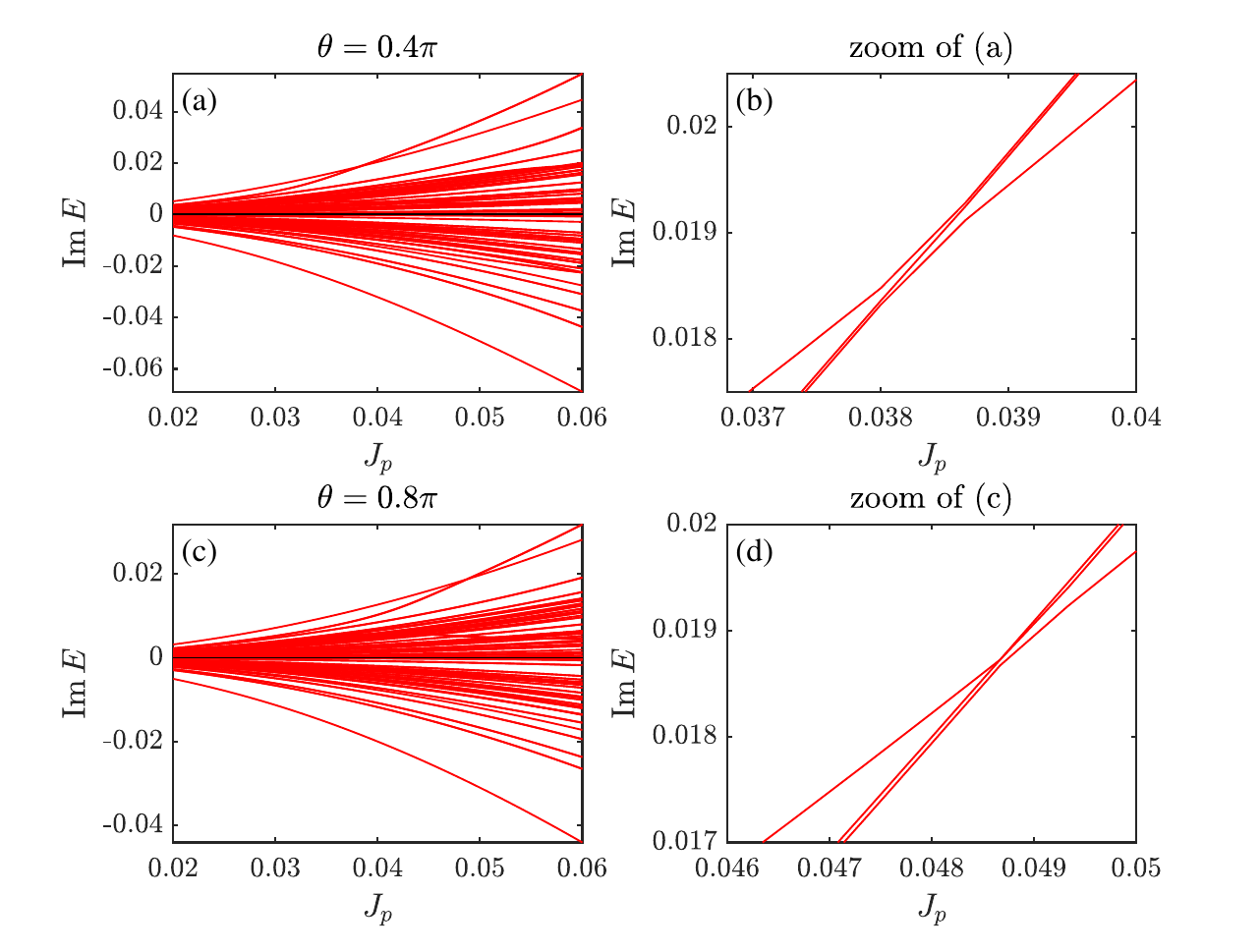}
\caption{\label{Around005}
(a) Imaginary energy levels around the crossing point for $\theta=0.4\pi$. (b) Zoomed-in view of (a) near the crossing point. (c), (d) the same as (a),(b) but for $\theta=0.8\pi$. We observe two nearly degenerate states. Parameters are $U=16,\mu=4,L=10$,  \( J e^{\alpha} = 0.5 \),\( J e^{-\alpha} = 1 \).
}  
\end{figure}

As discussed in the main text, we emphasize that the jump of the maximal imaginary energy is tied to a \emph{two-fold degeneracy} of the maximal $\mathrm{Im}\,E$ for anyons. To give a clear view,
we plot the imaginary energy levels versus $J_p$ for two anyonic cases in Fig. ~\ref{Around005}. As can be seen in the zoomed-in view near the crossing of $\max \mathrm{Im}\,E$, the maximal imaginary energy becomes nearly two-fold degenerate (with a small finite-size spacing in between) thereafter, which leads to the strong oscillation of the long-time quench dynamics in Fig.~\ref{fig:Quench}(a) of the main text.

\begin{figure}[htbp]
\centering
\includegraphics[height=4.5cm]{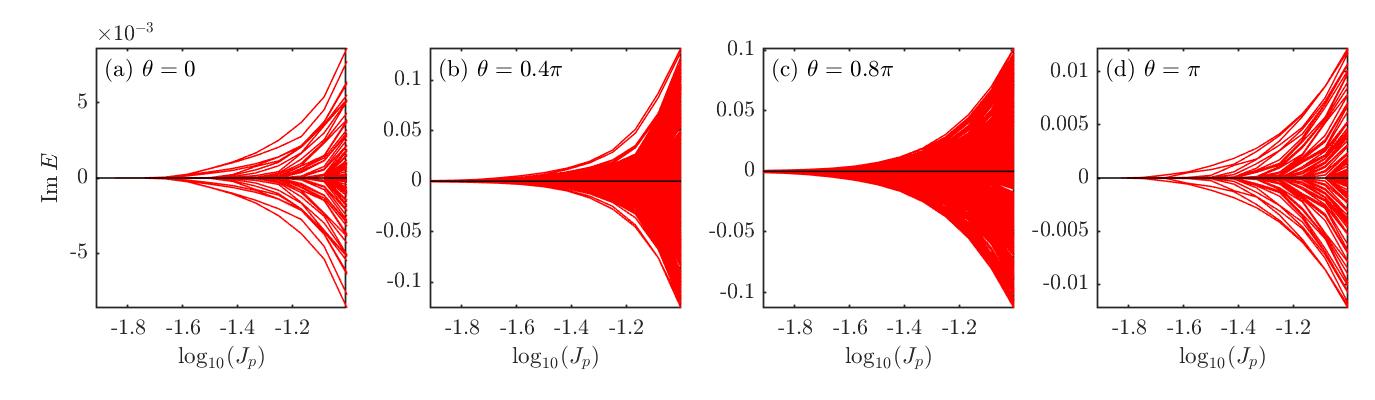}
\caption{\label{PTBroken_N3} Imaginary energy spectrum for \(N=3\), illustrating the effect of AIST. The results reveal the statistical angle as a powerful control knob for the non-Hermitian spectrum. While the differences between \(\theta=0.4\pi,\,0.8\pi\) and \(\theta=0,\,\pi\) are mainly quantitative, varying the statistical angle markedly reshapes the spectral distribution. In particular, the anyonic cases exhibit a lower critical coupling for the onset of complex eigenvalues, together with a broader and more densely populated distribution of imaginary parts than the non-anyonic cases \(\theta=0,\pi\). Other parameters are the same as Fig.~\ref{Around005}.   
}  
\end{figure}

It should be noted that the pseudo-Hermitian symmetry-breaking behavior for $\theta = 0$ and $\pi$ is quantitatively different from that for $\theta = 0.4\pi$ and $0.8\pi$, as shown in Fig.~\ref{PTBroken_N3} for $N=3$.
 Specifically, for sufficiently small (large) coupling, the spectrum remains real (complex) within the numerical resolution for all values of $\theta$. The key difference is that the number of growing and decaying states is significantly larger for $\theta$ away from $\theta=0$ and $\theta=\pi$. 

\section{Quench from the band with maximum imaginary eigenvalues}
\label{sec:sm-quench}

\subsection{Understanding the dynamical stability through eigenstate decomposition}

In order to understand the stability of anyons and the oscillations of bosons for $t<2000$, we evaluate the population on each eigenstate of the post-quench Hamiltonian. A simple calculation yields
\begin{align}
|\psi(t)\rangle 
&= e^{-iHt}|\psi(0)\rangle  \nonumber \\
&= \sum_n e^{-iHt}|\psi^R_{n}\rangle \langle \psi^L_{n}|\psi(0)\rangle \nonumber \\
&= \sum_n e^{-iE_n t}|\psi^R_{n}\rangle \langle \psi^L_{n}|\psi(0)\rangle \nonumber \\
&= \sum_n e^{-iE_n t}\langle \psi^L_{n}|\psi(0)\rangle\,|\psi^R_{n}\rangle \nonumber \\
&= \sum_n c_n(t)\,|\psi^R_{n}\rangle , \label{eq:decomposition}
\end{align}
where
\begin{equation}
c_n(t) = e^{-iE_n t}\langle \psi^L_{n}|\psi(0)\rangle 
       = \langle \psi^L_{n}|\psi(t)\rangle .
\end{equation}
The superscript $L$ and $R$ denote the left and right eigenstates of Hamiltonian $H$, respectively.

\begin{figure}[htbp]
\centering
\includegraphics[width=1\linewidth]{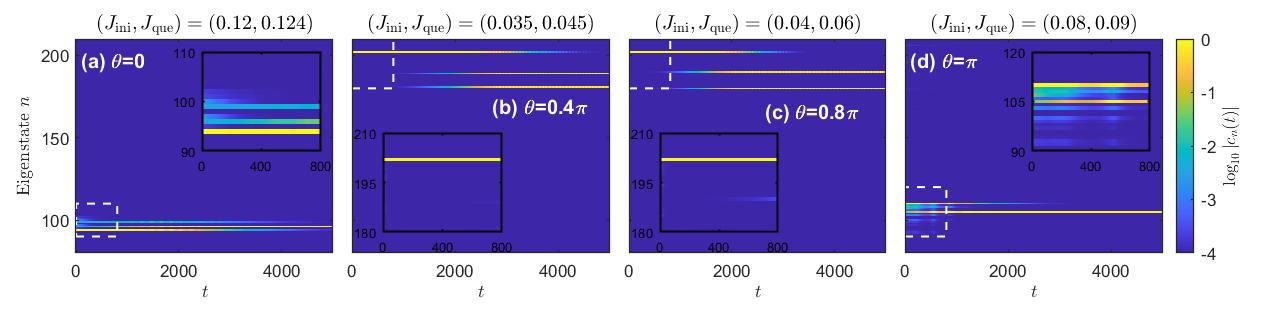}
\caption{\label{fig:DecomposeEigen}Decomposition into the eigenstates of the post-quench Hamiltonian for different values of $\theta$. Insets show zoomed-in views of the corresponding panels. Parameters: $L=10, N=2, U=16, \mu=4$.
}
\end{figure}

For bosons and pseudofermions, several eigenstates contribute within the time window $0$-$2000$, as shown in Fig.~\ref{fig:DecomposeEigen}. Since these eigenstates either decay or grow during the evolution, oscillations appear in the function $P(t)$. In contrast, for anyons with $\theta=0.4\pi$ and $0.8\pi$, only a single eigenstate dominates within the same time window, which accounts for their stability. However, beyond $t\approx2000$, bosons and pseudofermions are dominated by a single state with the largest imaginary component, while for anyons two nearly degenerate eigenstates emerge, leading to strong oscillations.

To intuitively understand the transition of $P(t)$, we use a two-state approximation. Suppose that, during a given stage of the dynamics, the state can be approximated by two relevant eigenstates of the post-quench Hamiltonian,
\begin{equation}
|\psi(t)\rangle \approx 
c_a e^{-iE_a t} |\psi_{E_a}\rangle +
c_b e^{-iE_b t} |\psi_{E_b}\rangle .
\end{equation}
For the anyonic quenches considered in the main text, the initially
dominant post-quench state corresponds to $E_2$ in the ordering
$\mathrm{Im}E_0\geq \mathrm{Im}E_1\geq \mathrm{Im}E_2\geq\cdots$,
while the growing competing state is one of the two top states
$E_0$ or $E_1$.
Here we keep the generic labels $E_a$ and $E_b$ because the following
two-state estimate is intended only to characterize the crossover
between the initially dominant state and a representative faster-growing
state. The initial state has a larger overlap with $|\psi_{E_a}\rangle$, corresponding to the smaller $\mathrm{Im}E$, and the dynamics will evolve toward $|\psi_{E_b}\rangle$ provided that $\mathrm{Im}E_b > \mathrm{Im}E_a$.
The corresponding time-dependent coefficients are
\begin{equation}
c_a(t) = c_a e^{-iE_a t}, \quad
c_b(t) = c_b e^{-iE_b t}.
\end{equation}
The ratio of their amplitudes evolves as
\begin{equation}
\frac{|c_a(t)|}{|c_b(t)|} 
= \frac{|c_a|}{|c_b|}\,
e^{(\mathrm{Im}E_a - \mathrm{Im}E_b)t}.
\end{equation}
Assuming $|c_a|/|c_b| \approx 10^{7}$ and $\mathrm{Im}E_b - \mathrm{Im}E_a \sim 4\times 10^{-3}$, the crossover time is estimated as
\begin{equation}
t = \frac{\ln(|c_a/c_b|)}{\mathrm{Im}E_b - \mathrm{Im}E_a} 
\sim 10^3.
\end{equation}

\begin{figure}[htbp]
\centering
\includegraphics[width=1\linewidth]{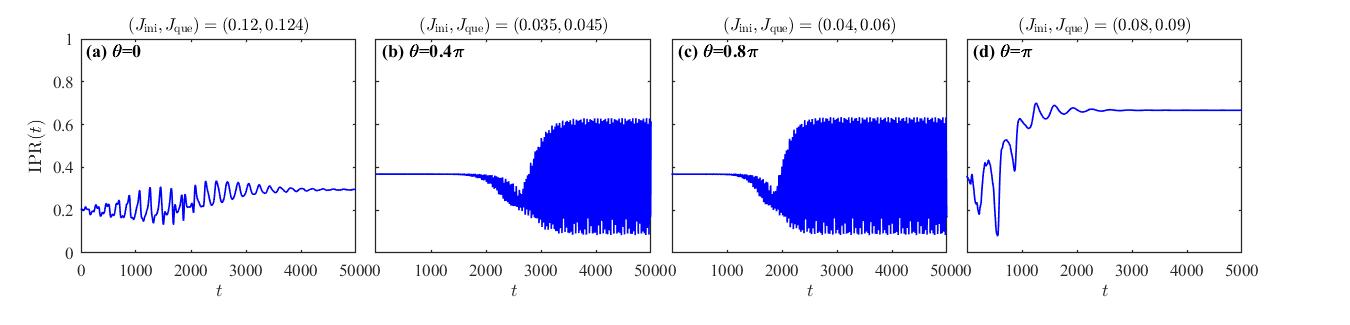}
\caption{\label{fig:IPRFock} $IPR$ by decomposing $|\psi(t)\rangle$ into Fock basis for different values of $\theta$.  $L=10,N=2,U=16,\mu=4$. 
}
\end{figure}

To characterize how broadly the evolving state spreads over the Fock basis, we define the inverse participation ratio (IPR)

\begin{equation}
\mathrm{IPR} = \sum_n |\langle n|\psi(t)\rangle|^4 .
\end{equation}
As shown in Fig.~\ref{fig:IPRFock}, the lower IPR values indicate that the wave function is distributed over a larger number of Fock states, while its oscillations reflect a time-dependent redistribution among these states.

\subsection{Density of states in imaginary energy for the post-quench Hamiltonian}

\begin{figure}[htbp]
\centering
\includegraphics[width=0.8\linewidth]{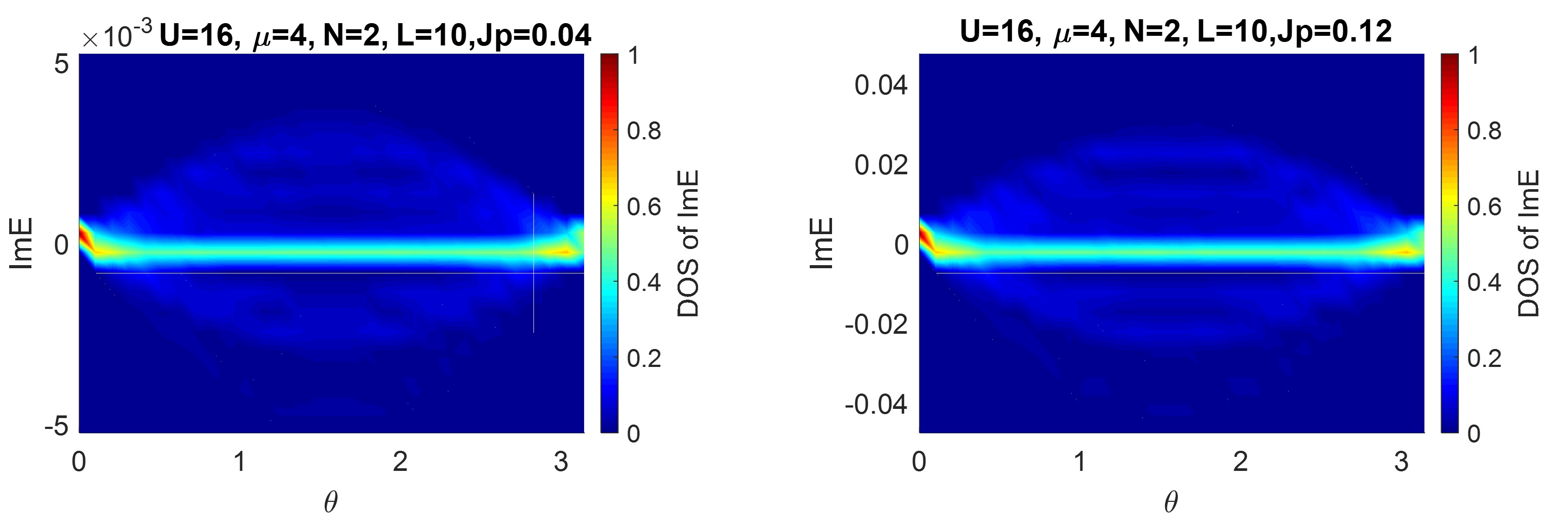}
\caption{\label{fig:DOSImE} Density of states (DOS) of imaginary energy of the post-quench Hamiltonian for two different coupling strengths.  $L=10,N=2,U=16,\mu=4$.  
}
\end{figure}

In this subsection, we present the density of states (DOS) in the imaginary parts of the eigenenergies of the post-quench Hamiltonian, as shown in Fig.~\ref{fig:DOSImE}. The energy spectrum is asymmetric with respect to $\mathrm{Im}E=0$, which arises from the breaking of pseudo-Hermitian symmetry. For bosons and pseudofermions, the imaginary part of the energy remains very close to zero.

\begingroup
\color{black}

\section{Quantitative analysis of imaginary-energy spacings}
\label{sec:sm_imag_spacing}

In this section we quantify the imaginary-energy spacings underlying the
spectral interpretation of the main-text Fig.~\ref{anyonPT} and the quench dynamics of
the main-text Fig.~\ref{fig:Quench}.  For each value of $J_p$ and statistical angle
$\theta$, we sort the imaginary parts of all eigenenergies in descending
order,
\begin{equation}
    \Im E_0 \geq \Im E_1 \geq \Im E_2 \geq \cdots .
\end{equation}
We then compute the adjacent spacings
\begin{equation}
    \delta_n = \Im E_n-\Im E_{n+1}.
\end{equation}
Since different physical statements refer to different parts of the spectrum,
we distinguish three quantities:
\begin{align}
    \delta_{\rm typ} &= {\rm median}_n\, \delta_n ,
    \label{eq:sm_delta_typ}\\
    \delta_{01} &= \Im E_0-\Im E_1 ,
    \label{eq:sm_delta01}\\
    \Delta_{\rm sep} &= \Im E_1-\Im E_2 .
    \label{eq:sm_delta_sep}
\end{align}
Here $\delta_{\rm typ}$ characterizes a typical full-spectrum spacing,
$\delta_{01}$ measures the splitting of the two states with largest imaginary
energies, and $\Delta_{\rm sep}$ measures the separation between this top pair
and the remaining spectrum.

\begin{figure}[t]
    \centering
    \includegraphics[width=\linewidth]{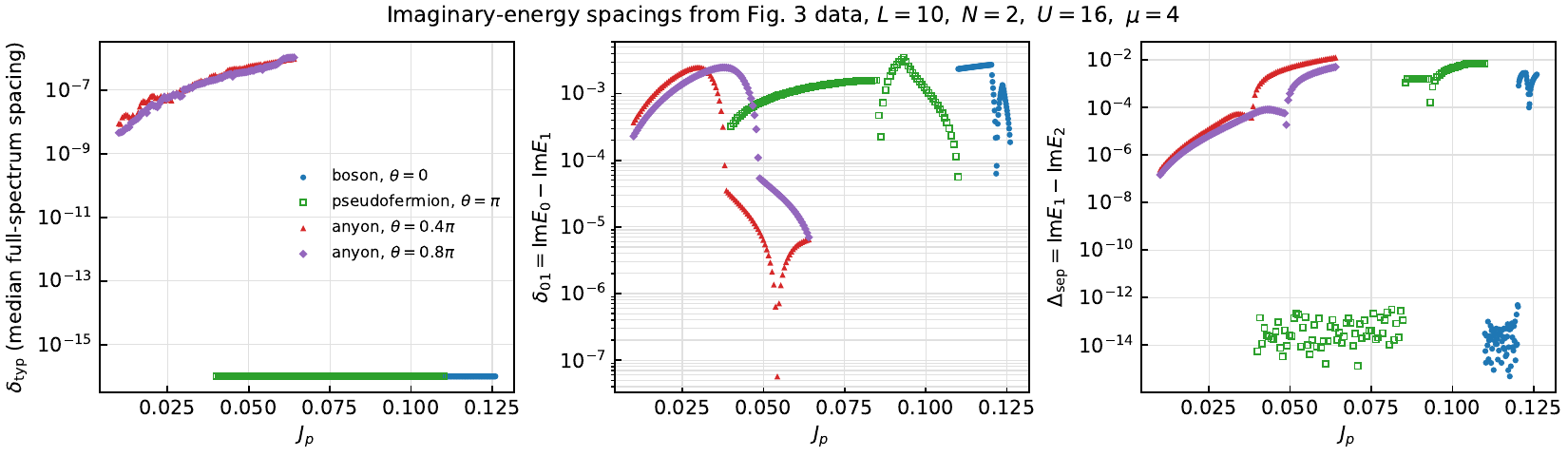}
    \caption{
    Quantitative comparison of imaginary-energy spacings for the spectra in
    the $J_p$ scans corresponding to the main-text Fig.~\ref{anyonPT}.  The three panels
    show the full-spectrum typical
    spacing $\delta_{\rm typ}$, the top-pair splitting $\delta_{01}$, and the
    separation $\Delta_{\rm sep}$ between the second and third largest
    imaginary energies.  The data are obtained for
    $N=2$, $U=16$, $\mu=4$, and $L=10$.
    }
    \label{fig:sm_imag_spacing_gaps}
\end{figure}

The post-quench gap hierarchy relevant to the dynamics is summarized in
Table~\ref{tab:sm_imag_spacing}.
For bosons and pseudofermions, most eigenenergies remain real or nearly real
in the parameter window of the main-text Fig.~\ref{anyonPT}.  Therefore the median of all
adjacent imaginary-energy spacings is close to numerical zero.  If these
near-zero spacings are removed, the typical nonzero spacing is instead of
order $10^{-3}$.  In contrast, for anyons with $\theta=0.4\pi$ and
$0.8\pi$, a broad set of states acquire nonzero imaginary energies, producing
a much denser nonzero imaginary spectrum with typical spacing of order
$10^{-7}$.

\begin{table}[t]
\caption{
Top-pair gap hierarchy at the post-quench parameters used in the main-text
Fig.~\ref{fig:Quench}.  Because the spectral data are evaluated on a discrete $J_p$ grid,
the table lists the nearest scanned value of $J_p$.
}
\label{tab:sm_imag_spacing}
\begin{ruledtabular}
\begin{tabular}{ccccc}
statistics
& $J_p$
& $\delta_{01}$
& $\Delta_{\rm sep}$
& $\Delta_{\rm sep}/\delta_{01}$
\\
\hline
boson, $\theta=0$
& $0.12392$
& $1.19\times 10^{-3}$
& $3.58\times 10^{-4}$
& $3.02\times 10^{-1}$
\\
pseudofermion, $\theta=\pi$
& $0.08970$
& $2.15\times 10^{-3}$
& $1.58\times 10^{-3}$
& $7.33\times 10^{-1}$
\\
anyon, $\theta=0.4\pi$
& $0.04510$
& $1.59\times 10^{-5}$
& $2.78\times 10^{-3}$
& $1.75\times 10^{2}$
\\
anyon, $\theta=0.8\pi$
& $0.06022$
& $1.56\times 10^{-5}$
& $3.71\times 10^{-3}$
& $2.38\times 10^{2}$
\end{tabular}
\end{ruledtabular}
\end{table}

These results clarify the gap terminology used in the main text.  The
anyon-induced stabilization is not caused by a uniformly enlarged
full-spectrum imaginary-energy spacing.  Rather, in the anyonic spectra the
two dominant states form a nearly degenerate top pair with a small
$\delta_{01}$, while this pair is separated from the rest of the spectrum by
a much larger $\Delta_{\rm sep}$.  For the quench parameters used in the
main-text Fig.~\ref{fig:Quench}, the post-quench anyonic spectra have
$\delta_{01}\sim 10^{-5}$ and $\Delta_{\rm sep}\sim 10^{-3}$, so that
$\Delta_{\rm sep}/\delta_{01}\sim 10^2$.  This explains why the
two-state description of the short-time dynamics is accurate for anyons,
whereas bosonic and pseudofermionic quenches involve stronger contributions
from other spectral components.

Finally, we note that only two system sizes are currently available for this
specific spacing analysis.  They are insufficient to distinguish a power-law
from an exponential finite-size scaling of $\delta_{01}$ or
$\Delta_{\rm sep}$.  The conclusions above therefore concern the gap
hierarchy at the system sizes and parameters used in the main-text Fig.~\ref{anyonPT} and
the corresponding quench calculations of the main-text Fig.~\ref{fig:Quench}, rather than an
asymptotic scaling law.

\endgroup

\section{Testing the novel dynamical stability under different parameters}
\label{sec:sm-stability}

\subsection{Test for the stability for different quench parameters}

\begin{figure}[htbp]
\centering
\includegraphics[width=0.9\linewidth]{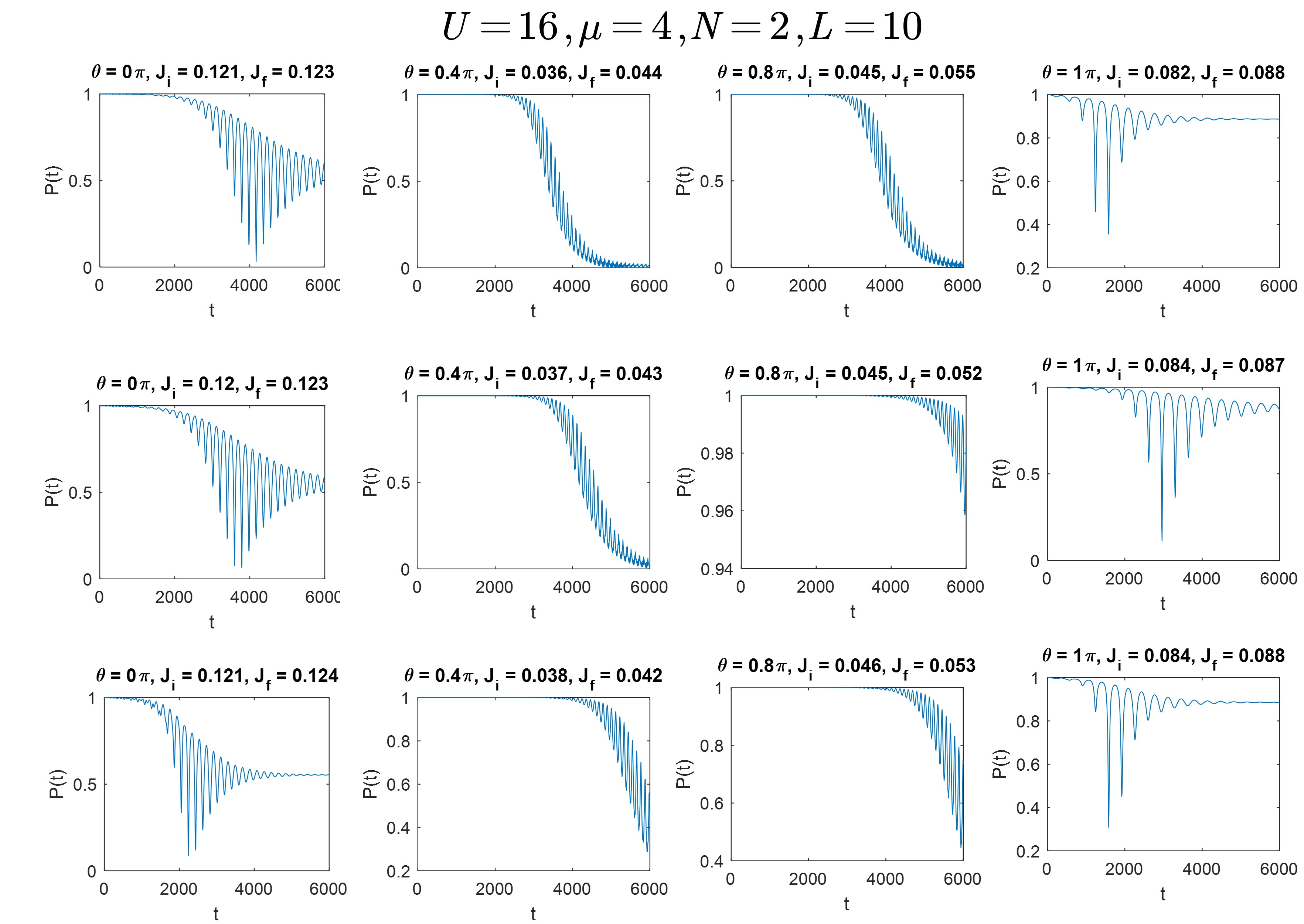}
\caption{\label{fig:U16mu4MoreJiJf} Different pairs of $J_{\rm ini}=J_i$ and $J_{\rm que}=J_f$ for various $\theta$. Parameters: $L=10, N=2, U=16, \mu=4$.}
\end{figure}

In this subsection, we calculate additional data for different quench values of $J_{\rm ini}=J_i$ and $J_{\rm que}=J_f$. The results are shown in Fig.~\ref{fig:U16mu4MoreJiJf}, which indicate that, under the same quench protocol, the dynamics of anyons is consistently more stable than that of bosons and pseudofermions. 

\subsection{Test for the stability of other $\theta$}

\begin{figure}[htbp]
\centering
\includegraphics[width=0.6\linewidth]{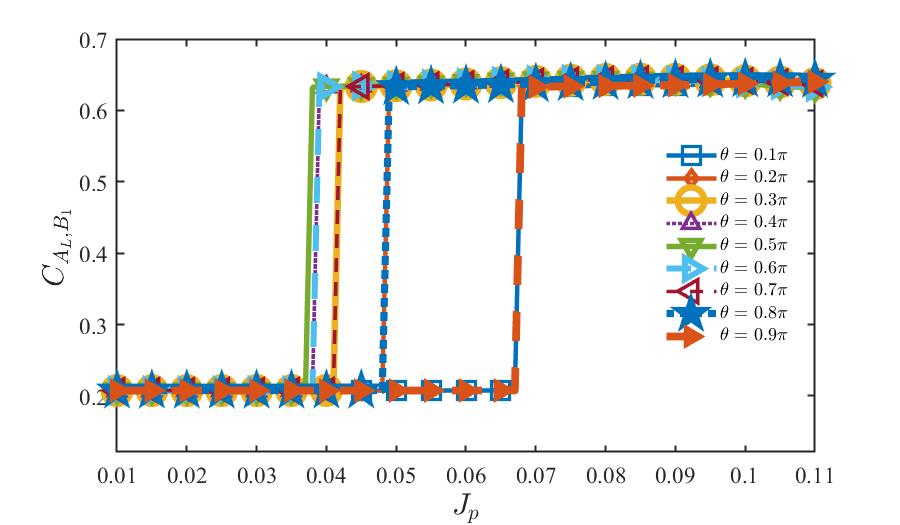}
\caption{\label{fig:Theta_CLB1A} The jump in the edge correlation function. Edge correlation function   $C_{AL,B1}=\langle \psi |\hat{n}_{A,L}\hat{n}_{B,1}|\psi \rangle$ . $|\psi \rangle$ is the eigenstate with maximum imaginary energy for a specific $J_p$. Parameters: $L=10, N=2, U=16, \mu=4$. }
\end{figure}

First, we calculate the edge correlation function for different statistical angles $\theta$, and the results are shown in Fig.~\ref{fig:Theta_CLB1A}. The results indicate that the boundary correlation is sensitive to all values of $\theta$.

\begin{figure}[htbp]
\centering
\includegraphics[width=0.8\linewidth]{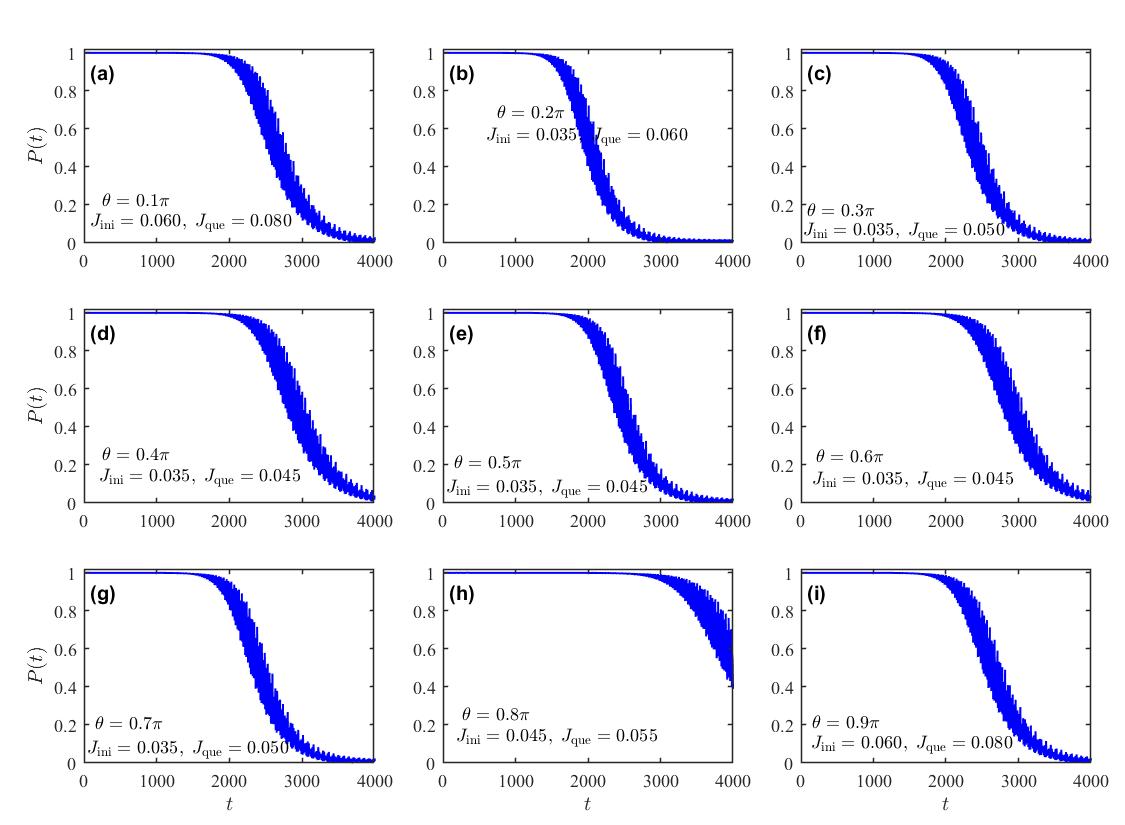}
\caption{\label{fig:Theta_quench} The quench dynamics around the jump point are always stable at short time. Parameters: $L=10, N=2, U=16, \mu=4$.
}
\end{figure}

Then, we quench the system from a smaller $J_p$ ($J_{\rm ini}$) to a larger $J_p$ ($J_{\rm que}$), as in the main text. The results are shown in Fig.~\ref{fig:Theta_quench}. These results indicate that the quench dynamics remains stable at short times, regardless of the choice of quench parameters.

\begingroup
\color{black}

\section{Finite-size diagnostics of the AIST}
\label{app:aist-finite-size-diagnostics}

We provide complementary finite-size diagnostics of the AIST, focusing on the
entanglement-based critical coupling, the weak-coupling spectral
complexification, and the strong-coupling saturation regime.

\subsection{Entanglement-based critical coupling}
\label{app:entropy-response-scaling}

We extract the finite-size entanglement-based critical coupling \(J_S^c(L)\)
from the first internal maximum of the entropy susceptibility
\(\chi_S(L,J_p)\). This prescription excludes the large-\(J_p\) saturation
regime and isolates the critical entropy response. For $\theta=0.4\pi$, the AIST, characterized by collective spectral
complexification, yields a stable sequence of \(J_S^c(L)\) that
systematically shifts toward weaker coupling with increasing \(L\). In
contrast, the \(\theta=0\) extrema are nonmonotonic and do not define a stable
entanglement-based critical coupling.

\paragraph*{Analytic origin of the exponential drift.}
The observed exponential drift can be understood from the relative
non-reciprocal localization length of the two decoupled legs. At
\(J_p=0\), each leg is a Hatano--Nelson chain up to onsite and interaction
terms. Under open boundary conditions, a non-unitary similarity
transformation removes the asymmetric hopping on each leg,
\begin{equation}
    \hat H_\sigma
    =
    S_\sigma \hat h_\sigma S_\sigma^{-1},
    \qquad
    S_\sigma
    =
    \exp\!\left[
    \kappa_\sigma\sum_j x_j \hat n_{j,\sigma}
    \right],
    \label{eq:skin-gauge-transformation}
\end{equation}
where \(x_j\) is measured from the center of a leg. For right and left
hopping amplitudes \(J_R^\sigma\) and \(J_L^\sigma\),
\begin{equation}
    \kappa_\sigma
    =
    \frac{1}{2}\log\left|
    \frac{J_R^\sigma}{J_L^\sigma}
    \right|.
\end{equation}
For the parameters used here,
\(J_R^A=1\), \(J_L^A=0.5\), \(J_R^B=0.5\), and \(J_L^B=1\), giving
\(\kappa_A=-\kappa_B=\frac{1}{2}\log2\). Thus the two legs exhibit
opposite skin localization.

We next examine how the inter-leg coupling transforms in this gauged
frame. Before the transformation, write
\begin{equation}
    \hat V_{AB}
    =
    \sum_j
    \left(
    \hat W_j+\hat W_j^\dagger
    \right),
    \qquad
    \hat W_j=\hat a_{j,A}^\dagger \hat a_{j,B},
\end{equation}
where possible anyonic occupation-dependent phase factors are included
in \(\hat W_j\). Since
\begin{equation}
    S_\sigma^{-1}\hat a_{j,\sigma}S_\sigma
    =
    e^{-\kappa_\sigma x_j}\hat a_{j,\sigma},
    \qquad
    S_\sigma^{-1}\hat a_{j,\sigma}^\dagger S_\sigma
    =
    e^{\kappa_\sigma x_j}\hat a_{j,\sigma}^\dagger,
\end{equation}
the transfer operators acquire a relative skin factor,
\begin{equation}
    \hat W_j
    \rightarrow
    e^{(\kappa_A-\kappa_B)x_j}\hat W_j,
    \qquad
    \hat W_j^\dagger
    \rightarrow
    e^{-(\kappa_A-\kappa_B)x_j}\hat W_j^\dagger,
\end{equation}
up to a convention-dependent sign in the exponent. The sign does not
affect the finite-size scaling because the largest matrix elements are
controlled by the magnitude of the relative factor. Since \(x_j\) is
measured from the center, \(|x_j|\le (L-1)/2\), and therefore the maximal
spatial envelope of the transformed inter-leg coupling scales as
\begin{equation}
    J_p
    \max_j
    e^{|\kappa_A-\kappa_B||x_j|}
    \sim
    J_p e^{|\kappa_A-\kappa_B|L/2}
    =
    J_p e^{\kappa_{\rm rel}L},
\end{equation}
where
\begin{equation}
    \kappa_{\rm rel}
    =
    \frac{|\kappa_A-\kappa_B|}{2}
    =
    \frac{1}{4}
    \left|
    \log\left|
    \frac{J_R^A J_L^B}{J_L^A J_R^B}
    \right|
    \right|.
    \label{eq:relative-skin-exponent}
\end{equation}
For the present parameters,
\begin{equation}
    \kappa_{\rm rel}
    =
    \frac{1}{2}\log2
    \simeq 0.347.
\end{equation}
This motivates the skin-dressed coupling
\begin{equation}
    g_L
    =
    J_p e^{\kappa_{\rm rel}L}
    \label{eq:entropy-scaling-variable}
\end{equation}
as the natural finite-size scaling variable.

We now make this argument more explicit at the level of eigenvector
hybridization. Away from exact degeneracies, and assuming that the
relevant eigenvalues of the decoupled Hamiltonian are diagonalizable,
standard biorthogonal perturbation theory gives, for a right eigenstate
\(\lvert r_n^{(0)}\rangle\),
\begin{equation}
    \lvert r_n(J_p)\rangle
    =
    \lvert r_n^{(0)}\rangle
    +
    \sum_{m\neq n}
    c_{mn}\lvert r_m^{(0)}\rangle
    +
    O(J_p^2),
\end{equation}
with
\begin{equation}
    c_{mn}
    =
    \frac{
    J_p
    \langle l_m^{(0)}|
    \hat V_{AB}
    |r_n^{(0)}\rangle
    }{
    E_n^{(0)}-E_m^{(0)}
    }.
\end{equation}
The gauge transformation shows that the spatial envelope of the relevant
inter-leg matrix elements can contain the factor
\(e^{\kappa_{\rm rel}L}\). For the states that govern the collective
crossover, we therefore write the leading scaling as
\begin{equation}
    \langle l_m^{(0)}|
    \hat V_{AB}
    |r_n^{(0)}\rangle
    \sim
    e^{\kappa_{\rm rel}L}
    \widetilde M_{mn}(L),
\end{equation}
where \(\widetilde M_{mn}(L)\) contains local overlaps, anyonic phase
factors, and residual finite-size dependence. This step is a scaling
assumption rather than an exact identity: it requires that the relevant
overlaps are not exponentially suppressed relative to the maximal
skin-enhanced envelope. Hence
\begin{equation}
    c_{mn}
    \sim
    \frac{
    g_L\,\widetilde M_{mn}(L)
    }{
    E_n^{(0)}-E_m^{(0)}
    }.
\end{equation}

The essential assumption is that, for the states controlling the
crossover, neither the reduced matrix elements
\(\widetilde M_{mn}(L)\) nor the inverse level spacings
\(1/|E_n^{(0)}-E_m^{(0)}|\) introduce an additional exponential scale
with system size. Equivalently, their finite-size dependence is assumed
to be at most subexponential. Under this condition, the opposite skin
localizations of the two legs provide the leading exponential
finite-size factor in the weak-to-intermediate coupling regime, and the
inter-leg hybridization is controlled by \(g_L\).

The same conclusion follows from a local two-state non-Hermitian
resonance. Consider
\begin{equation}
    H_{\rm couple}
    =
    \begin{pmatrix}
        E_1 & g_L\widetilde M \\
        g_L\widetilde M' & E_2
    \end{pmatrix}.
\end{equation}
Its two eigenvalues are
\begin{equation}
    \lambda_\pm
    =
    \frac{E_1+E_2}{2}
    \pm
    \sqrt{
        \frac{(E_1-E_2)^2}{4}
        +
        g_L^2\widetilde M\widetilde M'
    }.
\end{equation}
Thus the crossover to appreciable hybridization occurs when
\begin{equation}
    g_L^2
    |\widetilde M\widetilde M'|
    \sim
    |E_1-E_2|^2.
\end{equation}
If \(\widetilde M\), \(\widetilde M'\), and \(E_1-E_2\) carry no
additional exponential dependence on \(L\), this criterion again gives
\(g_L=O(1)\).

Motivated by this hybridization scale, we assume that the collective averaged
leg-entanglement (defined in Eq.~14 of End Matter) in the weak-to-intermediate coupling regime
obeys the single-parameter scaling form
\begin{equation}
    \Delta\overline{S_A}(L,J_p)
    \simeq
    F(g_L),
\end{equation}
up to algebraic and finite-size corrections. We emphasize that the
non-unitary similarity transformation is used here to identify the
relevant hybridization scale; the entanglement entropy itself is not
assumed to be invariant under this transformation. Rather, the scaling
form above is a physical scaling hypothesis for the entropy response in
the original frame.

With
\begin{equation}
    \chi_S(L,J_p)
    =
    \frac{
    \partial\Delta\overline{S_A}
    }{
    \partial\log J_p
    },
\end{equation}
the single-parameter scaling form gives
\begin{equation}
    \chi_S(L,J_p)
    \simeq
    g_L F'(g_L).
\end{equation}
If the first internal maximum of \(\chi_S\) occurs at an
\(L\)-independent value \(g_L=g_c\), then
\begin{equation}
    J_S^c(L)
    \simeq
    g_c e^{-\kappa_{\rm rel}L}.
    \label{eq:entropy-critical-exponential}
\end{equation}
Therefore the relative skin exponent predicts the exponential finite-size
drift of the entanglement-based critical coupling.

For the parameters considered here,
\(\kappa_{\rm rel}=(\log2)/2\simeq0.347\), while the numerical fits give
decay constants \(0.336\) for \(\theta=0.4\pi\) and \(0.329\) for
\(\theta=0.8\pi\). Their proximity to \((\log2)/2\) supports the
interpretation that the observed drift is governed primarily by the
relative skin localization. The remaining discrepancy is naturally
attributed to finite-size corrections, the finite \(J_p\) grid, and the
operational definition of \(J_S^c(L)\) through the susceptibility
maximum.

The statistical angle modifies the complex phases and prefactors of the
inter-leg matrix elements but, for generic \(\theta\neq0,\pi\), does not
change the relative skin exponent in
Eq.~\eqref{eq:relative-skin-exponent}. This explains why the two anyonic
angles exhibit nearly the same exponential drift. At special statistical
angles, however, symmetry or destructive-interference effects may
suppress the leading matrix elements, in which case subleading processes
can control the finite-size response.

\begin{figure}[!t]
    \centering
    \includegraphics[width=0.4\columnwidth]{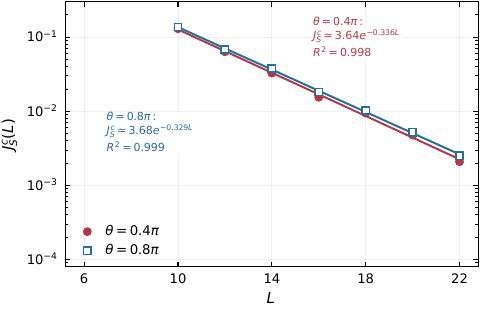}
     \caption{Finite-size drift of the averaged-entropy response scale
    \(J_S^c(L)\), extracted from the first internal maximum of
    \(\chi_S(L,J_p)\).  The \(\theta=0.4\pi\) data follow
    \(J_S^c(L)\simeq3.64e^{-0.336L}\) with \(R^2\simeq0.998\), while
    \(\theta=0.8\pi\) gives \(J_S^c(L)\simeq3.68e^{-0.329L}\) with
    \(R^2\simeq0.999\), showing that the collective entropy scale drifts
    toward weak coupling with increasing system size for both anyonic angles.}
    \label{fig:entropy-response-jstar-scaling}
\end{figure}

\begin{figure}[!t]
    \centering
    \includegraphics[width=0.6\columnwidth]{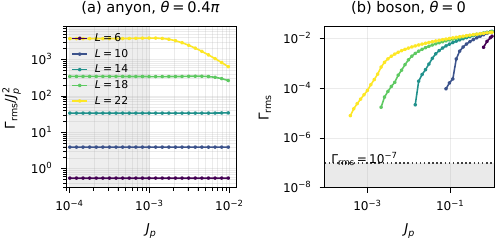}
    \caption{Weak-coupling check of the spectral-complexification scaling.
    (a) \(\Gamma_{\rm rms}/J_p^2\) for \(\theta=0.4\pi\); the nearly constant
    curves over \(10^{-4}\le J_p\le10^{-3}\) confirm
    \(\Gamma_{\rm rms}=C_L(\theta)J_p^2\).
    (b) Bosonic data after excluding the numerical-noise floor
    \(\Gamma_{\rm rms}\le10^{-7}\). The absence of a weak-coupling signal is
    consistent with \(C_L(0)=0\), while complexification appears only beyond
    the finite-size spectral threshold.}
    \label{fig:gamma-weak-coupling-validation}
\end{figure}
\subsection{Weak-coupling spectral complexification}
\label{app:weak-coupling-gamma}

We next derive the weak-coupling behavior of the spectral-complexification
scale in Fig.~\ref{fig:endmatter-spectral-entanglement} of the main text. We write
\begin{equation}
    \hat H_{\rm eff}(J_p)
    =
    \hat H_0+J_p\hat V_{AB},
    \qquad
    J_p\hat V_{AB}\equiv\hat H_{AB}^{\rm eff}.
\end{equation}
At \(J_p=0\), the particle numbers on the  two legs are separately
conserved. Let \(P_m\) project onto the sector with \(N_A=m\) and
\(N_B=N-m\). Since \(\hat V_{AB}\) transfers one particle between the two
legs,
\begin{equation}
    P_m\hat V_{AB}P_m=0 .
    \label{eq:inter-sublattice-selection-rule}
\end{equation}

For an isolated eigenvalue of \(\hat H_0\), we define the right and left
eigenvectors by
\begin{equation}
    \hat H_0|r_\alpha^{(m)}\rangle
    =
    E_\alpha^{(0,m)}|r_\alpha^{(m)}\rangle,
    \qquad
    \langle l_\alpha^{(m)}|\hat H_0
    =
    E_\alpha^{(0,m)}\langle l_\alpha^{(m)}|,
\end{equation}
with
\begin{equation}
    \langle l_\alpha^{(m)}|r_\beta^{(m')}\rangle
    =
    \delta_{\alpha\beta}\delta_{mm'}.
\end{equation}
Equation~\eqref{eq:inter-sublattice-selection-rule} immediately gives
\begin{equation}
    E_\alpha^{(1)}
    =
    \langle l_\alpha^{(m)}|
    \hat V_{AB}
    |r_\alpha^{(m)}\rangle
    =
    0.
\end{equation}
The leading correction is therefore second order,
\begin{equation}
    E_\alpha(J_p)
    =
    E_\alpha^{(0,m)}
    +
    J_p^2\eta_\alpha^{(m)}
    +
    O(J_p^4),
    \label{eq:second-order-energy}
\end{equation}
where
\begin{equation}
    \eta_\alpha^{(m)}
    =
    \sum_{m'=m\pm1}
    \sum_\beta
    \frac{
    \langle l_\alpha^{(m)}|\hat V_{AB}|r_\beta^{(m')}\rangle
    \langle l_\beta^{(m')}|\hat V_{AB}|r_\alpha^{(m)}\rangle
    }{
    E_\alpha^{(0,m)}-E_\beta^{(0,m')}
    } .
    \label{eq:eta-second-order}
\end{equation}
Hence
\begin{equation}
    {\rm Im}\,E_\alpha(J_p)
    =
    J_p^2\,{\rm Im}\,\eta_\alpha^{(m)}
    +
    O(J_p^4),
\end{equation}
and
\begin{equation}
    \begin{aligned}
    \Gamma_{\rm rms}(L,J_p)
    &=
    C_L(\theta)J_p^2+O(J_p^4),\\
    C_L(\theta)
    &=
    \left[
    \frac{1}{D(L)}
    \sum_\alpha
    \bigl({\rm Im}\,\eta_\alpha\bigr)^2
    \right]^{1/2}.
    \end{aligned}
    \label{eq:gamma-rms-weak-coupling}
\end{equation}
Whenever \(C_L(\theta)\neq0\),
\begin{equation}
    \frac{d\log\Gamma_{\rm rms}}{d\log J_p}
    =
    2+O(J_p^2).
\end{equation}

For anyons, the inter-leg hopping carries an occupation-dependent
Peierls phase,
\begin{equation}
    \begin{aligned}
    \hat a_{j,B}^\dagger\hat a_{j,A}
    &\rightarrow
    f_j(\theta)\hat b_{j,B}^\dagger\hat b_{j,A},\\
    f_j(\theta)
    &=
    \exp\!\left[
    i\theta\left(
    \sum_{k=j}^L\hat n_{k,A}
    +
    \sum_{k=1}^{j-1}\hat n_{k,B}
    \right)
    \right],
    \end{aligned}
\end{equation}
and similarly for the Hermitian-conjugate hopping. The products entering
Eq.~\eqref{eq:eta-second-order} are therefore generically complex, giving
\(C_L(\theta)\neq0\) for generic anyonic angle, including
\(\theta=0.4\pi\). This produces the quadratic weak-coupling opening
\begin{equation}
    \Gamma_{\rm rms}\propto J_p^2,
\end{equation}
characteristic of the perturbative onset of the AIST.

The bosonic limit \(\theta=0\) is qualitatively different. Both
\(\hat H_0\) and \(\hat V_{AB}\) can be represented by real matrices, so an
isolated real eigenvalue has a real perturbative expansion around \(J_p=0\).
In particular,
\begin{equation}
    {\rm Im}\,\eta_\alpha=0,
    \qquad
    C_L(0)=0 .
\end{equation}
Thus, within the analytic region below the finite-size spectral threshold,
\begin{equation}
    {\rm Im}\,E_\alpha(J_p)=0,
    \qquad
    \Gamma_{\rm rms}(L,J_p)=0.
\end{equation}
Complex eigenenergies appear only after the finite-size real-complex
transition. If \(J_c(L)\) denotes the corresponding spectral critical coupling,
a generic two-level normal form gives
\begin{equation}
    E_\pm(J_p)
    \simeq
    E_c\pm B_L\sqrt{J_p-J_c(L)},
\end{equation}
and therefore
\begin{equation}
    \Gamma_{\rm rms}(L,J_p)
    \sim
    A_L\sqrt{J_p-J_c(L)}
\end{equation}
immediately above the threshold. The apparent bosonic slope is therefore a
finite-size crossover slope rather than a weak-coupling perturbative exponent. These weak-coupling predictions are confirmed numerically in
Fig.~\ref{fig:gamma-weak-coupling-validation}.

\subsection{Strong-coupling saturation}

The large-\(J_p\) regime provides a consistency check but does not determine
\(J_S^c(L)\). Writing
\begin{equation}
    \hat H_{\rm eff}(J_p)
    =
    J_p\hat V_{AB}+\hat H_0
    =
    J_p\left(
    \hat V_{AB}
    +
    \lambda\hat H_0
    \right),
    \qquad
    \lambda=\frac{1}{J_p},
\end{equation}
shows that the overall factor \(J_p\) does not affect the right eigenvectors.
Away from exact degeneracies,
\begin{equation}
    |r_n(J_p)\rangle
    =
    |\varphi_n^{(0)}\rangle
    +
    \frac{1}{J_p}|\varphi_n^{(1)}\rangle
    +
    O(J_p^{-2}),
\end{equation}
so the averaged entropy approaches a finite limit,
\begin{equation}
    \overline{S_A}(L,J_p)
    =
    \overline{S_A}^{(\infty)}(L)
    +
    \frac{a_L}{J_p}
    +
    O(J_p^{-2}).
\end{equation}
Consequently,
\begin{equation}
    \chi_S(L,J_p)
    =
    \frac{\partial\overline{S_A}}{\partial\log J_p}
    =
    -\frac{a_L}{J_p}
    +
    O(J_p^{-2})
    \to0 .
\end{equation}
The large-\(J_p\) side is therefore a saturation crossover rather than the
critical regime used to extract \(J_S^c(L)\). The corresponding strong-coupling asymptotics are verified numerically in
Fig.~\ref{fig:strong-coupling-entropy-asymptotics}.

\begin{figure}[!t]
    \centering
    \includegraphics[width=0.6\columnwidth]{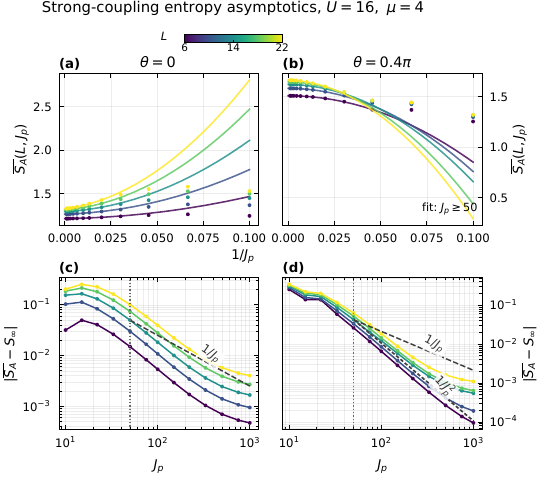}
    \caption{Strong-coupling saturation of the averaged entropy. The data are
    obtained from the scaled Hamiltonian
    \(\hat V_{AB}+J_p^{-1}\hat H_0\), which has the same right eigenvectors as
    \(\hat H_{\rm eff}(J_p)\) and is better conditioned at large \(J_p\).
    The upper panels show \(\overline{S_A}\) versus \(1/J_p\), with quadratic
    fits using \(J_p\ge50\); the lower panels show residuals from the fitted
    intercept \(S_\infty(L)\). Both the bosonic and anyonic cases approach
    finite strong-coupling entropies, confirming that this regime is a
    saturation crossover rather than the entropy critical regime.}
    \label{fig:strong-coupling-entropy-asymptotics}
\end{figure}

\section{Quantum-computing implementation methods}
\label{app:qc-details}

\emph{Hardware implementation of the dynamical states.---}
For the quantum-hardware implementation, we consider the normalized
post-quench states
\begin{equation}
|\psi(t)\rangle =
\frac{e^{-iH(J_{\mathrm{final}})t}|\psi_{\rm ini}\rangle}
{\sqrt{\langle\psi_{\rm ini}|
e^{+iH^\dagger(J_{\mathrm{final}})t}
e^{-iH(J_{\mathrm{final}})t}
|\psi_{\rm ini}\rangle}},
\end{equation}
where $|\psi_{\rm ini}\rangle$ is the right eigenstate of
$H(J_{\mathrm{ini}})$ with the largest imaginary eigenenergy.
The target amplitudes used for circuit compilation are obtained by
numerically evaluating this normalized evolution within the two particle case.

The numerical trajectory contains 81 time points up to $t=5000$.
For the hardware demonstration, we select eleven representative times,
\begin{equation}
i_t=0,8,16,\ldots,80,
\qquad
t=0,500,1000,\ldots,5000 .
\end{equation}
The bosonic reference calculation uses the same eleven times and the
same compact Fock-space encoding. In the present data set, the bosonic
hardware run focuses on the overlap observable $P(t)$.

\emph{Fock-space encoding.---}
The ladder considered in the quantum-hardware demonstration has
$L=10$ unit cells and two legs, corresponding to $2L=20$
physical modes. For two bosonic occupation quanta, the
fixed-particle-number Fock space contains
\begin{equation}
D=\binom{2L+N-1}{N}
=\binom{21}{2}
=210
\end{equation}
basis states. We therefore use a compact binary encoding requiring only
\begin{equation}
n_q=\lceil \log_2 D\rceil=8
\end{equation}
qubits.

If
\begin{equation}
|\psi\rangle
=
\sum_{\alpha=0}^{D-1}
\psi_\alpha
|\alpha\rangle_{\rm Fock},
\end{equation}
the corresponding encoded register state is
\begin{equation}
|\psi\rangle_{\rm enc}
=
\sum_{\alpha=0}^{D-1}
\psi_\alpha|\alpha\rangle
+
\sum_{\alpha=D}^{255}
0|\alpha\rangle .
\end{equation}
The computational-basis states with integer labels $\alpha\ge D$
constitute a padding sector. They are present in the eight-qubit
register but carry zero amplitude in the target encoded state.

This compact representation is particularly useful for few-particle
many-body problems because the required number of qubits scales with
the logarithm of the fixed-particle-number Hilbert-space dimension,
rather than with the total number of physical modes.

\emph{Variational circuits.---}
Each selected time-dependent state is implemented using a variational circuit $W_t$ satisfying
\begin{equation}
W_t|0\rangle^{\otimes8}
\simeq
|\psi(t)\rangle_{\rm enc}.
\end{equation}
A reference circuit $W_{\rm ref}$ prepares
$|\psi_{\rm ini}\rangle_{\rm enc}$.

The circuit ansatz consists of $R_yR_z$ rotations on every qubit
interleaved with a brick-wall CNOT pattern. The entangling layers
alternate between
\begin{equation}
(0,1),(2,3),(4,5),(6,7)
\end{equation}
and
\begin{equation}
(1,2),(3,4),(5,6).
\end{equation}
For the anyonic hardware run, ten entangling layers were used, giving
\begin{equation}
N_{\rm par}
=
2n_q(\mathrm{layers}+1)
=
176
\end{equation}
trainable rotation parameters for each circuit.
The corresponding logical block contains 35 CNOT
gates. The overlap circuit is deeper because it implements
$W_{\rm ref}^{\dagger}W_t$ before measurement.

The variational parameters are optimized with L-BFGS-B by maximizing
the preparation fidelity with the corresponding target encoded state.
For the anyonic run, all reference and time-dependent circuits reach
the target fidelity
\begin{equation}
F_{\rm target}=0.99,
\end{equation}
with achieved preparation fidelities ranging from
$0.990001$ to $0.990225$.
The resulting shallow circuits provide a hardware-efficient
realization of the time-dependent many-body states in the compact
encoded Hilbert space.

For the bosonic reference run, the overlap-only circuits use the same
eight-qubit encoding with a 14-layer ansatz.
The achieved preparation fidelities range from $0.967337$ to
$0.990260$. The later-time states reach the $0.99$ target, while
several early-time states remain slightly below this value.
We quantify the associated preparation and hardware effects by
comparing the numerical reference, noiseless circuit simulation,
raw hardware data, and M3+ZNE results.

\emph{Overlap observable.---}
The hardware data shown in Fig.~\ref{fig:qc-hardware-demo} of the \textbf{End Matter} use the
overlap observable $P(t)$. The corresponding circuit first prepares
the encoded time-dependent state, then applies the inverse reference
circuit,
\begin{equation}
|0\rangle^{\otimes8}
\xrightarrow{W_t}
W_t|0\rangle^{\otimes8}
\xrightarrow{W_{\rm ref}^{\dagger}}
W_{\rm ref}^{\dagger}
W_t|0\rangle^{\otimes8}.
\end{equation}
The probability of measuring the all-zero bit string is therefore
\begin{equation}
\Pr(0^8)
=
\left|
\langle 0|^{\otimes8}
W_{\rm ref}^{\dagger}
W_t
|0\rangle^{\otimes8}
\right|^2
\simeq
|\langle\psi_{\rm ini}|\psi(t)\rangle|^2,
\end{equation}
and the hardware overlap estimator is
\begin{equation}
P_{\rm hw}(t)
=
\sqrt{\Pr(0^8)}.
\end{equation}

This construction directly measures the overlap between the encoded
initial and time-dependent many-body states without requiring
ancilla-based tomography or reconstruction of the full wave function.
The data reported in Fig.~\ref{fig:qc-hardware-demo} of the \textbf{End Matter} are obtained
from this estimator for both the $\theta=0.4\pi$ anyonic problem and
the corresponding $\theta=0$ bosonic reference.

\emph{IBM Runtime implementation.---}
The anyonic demonstration was executed on
\texttt{ibm\_berlin} through the European-region IBM Quantum Platform.
The Runtime job identifier was
\texttt{d9cbln85f4ls73beb4bg}.
The overlap data in Fig.~\ref{fig:qc-hardware-demo} use one overlap circuit for each of the
eleven sampled times, with $20000$ shots per circuit.
The same job also contained density-readout circuits that are not
included in the plotted End Matter data.

The circuits were submitted using
\texttt{qiskit}=2.2.3 and
\texttt{qiskit-ibm-runtime}=0.44.0 through
\texttt{SamplerV2}.
Transpilation targeted \texttt{ibm\_berlin}, used optimization level
3, eight transpiler-seed trials per circuit, and base seed 20260713.
The selected anyonic overlap circuits contained approximately
70 two-qubit operations with circuit depth 112.
For example, the first overlap circuit mapped the eight logical qubits
to physical qubits
$17,16,15,14,13,12,11,10$
and contained 296 \texttt{rz} gates, 296 \texttt{sx} gates,
70 \texttt{cz} gates, and eight measurements.

The bosonic reference run was also executed on
\texttt{ibm\_berlin}, with $20000$ shots for each overlap circuit.
After transpilation, the selected bosonic overlap circuits contained
98 two-qubit gates and had depth 152.
For each of the eleven sampled times, an independent M3+ZNE job was
submitted using folded overlap circuits with noise scales
$s=1,3,5$.

\emph{Error mitigation.---}
The raw hardware data are affected predominantly by systematic
hardware noise rather than finite-shot uncertainty.
Measurement-error mitigation is performed using M3, followed by
zero-noise extrapolation for the overlap observable.

The noise scaling is generated by global unitary folding of the
transpiled overlap circuit,
\begin{equation}
U
\longrightarrow
U(U^\dagger U)^k,
\qquad
s=2k+1,
\end{equation}
with $s=1,3,5$.
M3 is applied independently at each noise scale, and the reported
M3+ZNE value is obtained from the linear intercept of
$P_{\rm M3}(s)$ as a function of $s$.

At the first time point, the scale-dependent M3 values are
\begin{equation}
P_{\rm M3}(s=1,3,5)
=
0.893230,\;
0.741843,\;
0.582674,
\end{equation}
which gives
\begin{equation}
P_{\rm M3+ZNE}
=
0.972166.
\end{equation}
Bootstrap error bars are obtained by multinomial resampling of the
measured counts.

For the bosonic reference run, the same M3+ZNE procedure gives close
agreement with the numerical overlap reference over the full time
series. Denoting the numerical reference by $P_{\rm ref}$ and the
mitigated hardware result by $P_{\rm hw}^{\rm mit}$, representative
values are
\begin{align}
P_{\rm ref}(0,500,1000)
&=
1.000000,\;
0.996505,\;
0.975285,
\\
P_{\rm hw}^{\rm mit}(0,500,1000)
&=
0.990705,\;
0.980569,\;
0.947631,
\end{align}
and
\begin{align}
P_{\rm ref}(3000,4000,5000)
&=
0.475908,\;
0.537664,\;
0.556395,
\\
P_{\rm hw}^{\rm mit}(3000,4000,5000)
&=
0.491919,\;
0.509792,\;
0.517894.
\end{align}

The residual difference between the hardware measurements and the
numerical reference is consistent with current NISQ-device
limitations. The overlap circuit is particularly noise-sensitive
because it contains both the time-dependent block
and the inverse reference-state block before the zero-string
measurement. Two-qubit gate errors, readout-assignment errors,
decoherence during circuit execution, and leakage into the padding
sector of the compact encoding all contribute.

Because the target-state preparation fidelities are already close to
$0.99$ and the bootstrap statistical uncertainties are small,
increasing the number of shots alone is not expected to remove the
dominant discrepancy. Further improvements can instead be obtained
through shallower circuits, more extensive
transpiler-seed searches, improved backend selection, and more
systematic error-mitigation procedures.

\emph{Boundary-correlator readout.---}
Although Fig.~\ref{fig:qc-hardware-demo} focuses on the overlap observable, the same compact
encoding also provides a direct route to measuring the boundary
correlator $C_{AL,B1}(t)$.
After preparing
$W_t|0\rangle^{\otimes8}$,
all qubits are measured in the computational basis.
For each valid bit string $\alpha<D$, the corresponding decoded Fock
occupation vector ${\bf n}^{(\alpha)}$ gives the estimator
\begin{equation}
C_{AL,B1}(t)
=
\frac{
\sum_{\alpha<D}
N_\alpha
n_{A,L}^{(\alpha)}
n_{B,1}^{(\alpha)}
}{
\sum_{\alpha<D}
N_\alpha
}.
\end{equation}
Shots in the padding sector $\alpha\ge D$ are recorded as leakage and
excluded from the normalized physical-sector average.
This density-readout circuit is shallower than the overlap circuit
because it does not require the inverse reference-state preparation.

\emph{Broader applicability of the implementation.---}
The hardware strategy used here is not specific to the present
anyon ladder. Its main ingredients---compact encoding of a
fixed-particle-number Hilbert space, shallow variational state
preparation, overlap measurements, and direct decoding of
occupation-basis observables---can be adapted to other few-particle
bosonic, fermionic, and anyonic lattice models.
The same framework is also applicable to non-Hermitian quench
problems in which directly compiling the full non-unitary generator
into a gate-by-gate hardware evolution would require substantially
deeper circuits.

In this sense, the present implementation provides a general
hardware-efficient route for probing many-body dynamical signatures
and comparing distinct particle statistics within experimentally
accessible circuit depths.

\endgroup

\makeatletter
\patchcmd{\endthebibliography}{\label{LastBibItem}}{\label{LastSuppBibItem}}{}{\PackageError{arxiv}{Could not separate bibliography labels}{}}
\patchcmd{\NAT@bibsetnum}{\ref{LastBibItem}}{\ref{LastSuppBibItem}}{}{\PackageError{arxiv}{Could not separate bibliography widths}{}}
\makeatother
%

\end{document}